\documentclass[aps,prc,twocolumn,superscriptaddress]{revtex4-1}

\usepackage{graphicx}
\usepackage{multirow}
\usepackage{subfigure}
\usepackage{float}
\usepackage{dcolumn}
\usepackage{bm}
\usepackage{dcolumn}
\usepackage{siunitx}
\newcolumntype{d}[1]{D{.}{.}{#1}}
\begin{document}


\title{$\beta$ and $\beta$-delayed neutron decay of the $N=82$ nucleus $^{131}_{~49}$In$_{82}$}


\author{R.~Dunlop}
\email[]{rdunlop@uoguelph.ca}
\affiliation{Department of Physics, University of Guelph, Guelph, Ontario N1G 2W1, Canada}

\author{C.~E.~Svensson}
\affiliation{Department of Physics, University of Guelph, Guelph, Ontario N1G 2W1, Canada}

\author{C.~Andreoiu}
\affiliation{Department of Chemistry, Simon Fraser University, Burnaby, British Columbia V5A 1S6, Canada}


\author{G.~C.~Ball}
\affiliation{TRIUMF, 4004 Wesbrook Mall, Vancouver, British Columbia V6T 2A3, Canada}

\author{N.~Bernier}
\affiliation{TRIUMF, 4004 Wesbrook Mall, Vancouver, British Columbia V6T 2A3, Canada}
\affiliation{Department of Physics and Astronomy, University of British Columbia, Vancouver, British Columbia V6T 1Z4, Canada}

\author{H.~Bidaman}
\affiliation{Department of Physics, University of Guelph, Guelph, Ontario N1G 2W1, Canada}

\author{V.~Bildstein}
\affiliation{Department of Physics, University of Guelph, Guelph, Ontario N1G 2W1, Canada}

\author{M.~Bowry}
\affiliation{TRIUMF, 4004 Wesbrook Mall, Vancouver, British Columbia V6T 2A3, Canada}

\author{D.~S.~Cross}
\affiliation{Department of Chemistry, Simon Fraser University, Burnaby, British Columbia V5A 1S6, Canada}

\author{I.~Dillmann}
\affiliation{TRIUMF, 4004 Wesbrook Mall, Vancouver, British Columbia V6T 2A3, Canada}
\affiliation{Department of Physics and Astronomy, University of Victoria, Victoria, British Columbia V8P 5C2, Canada}

\author{M.~R.~Dunlop}
\affiliation{Department of Physics, University of Guelph, Guelph, Ontario N1G 2W1, Canada}

\author{F.~H.~Garcia}
\affiliation{Department of Chemistry, Simon Fraser University, Burnaby, British Columbia V5A 1S6, Canada}

\author{A.~B.~Garnsworthy}
\affiliation{TRIUMF, 4004 Wesbrook Mall, Vancouver, British Columbia V6T 2A3, Canada}

\author{P.~E.~Garrett}
\affiliation{Department of Physics, University of Guelph, Guelph, Ontario N1G 2W1, Canada}

\author{G.~Hackman}
\affiliation{TRIUMF, 4004 Wesbrook Mall, Vancouver, British Columbia V6T 2A3, Canada}

\author{J.~Henderson}
\altaffiliation{Present Address: Lawrence Livermore National Laboratory, 7000 East Ave, Livermore, CA 94550, USA}
\affiliation{TRIUMF, 4004 Wesbrook Mall, Vancouver, British Columbia V6T 2A3, Canada}

\author{J.~Measures}
\affiliation{TRIUMF, 4004 Wesbrook Mall, Vancouver, British Columbia V6T 2A3, Canada}
\affiliation{Department of Physics, University of Surrey, Guildford GU2 7XH, United
Kingdom}

\author{D. M\"{u}cher}
\affiliation{Department of Physics, University of Guelph, Guelph, Ontario N1G 2W1, Canada}

\author{B.~Olaizola}
\affiliation{Department of Physics, University of Guelph, Guelph, Ontario N1G 2W1, Canada}
\affiliation{TRIUMF, 4004 Wesbrook Mall, Vancouver, British Columbia V6T 2A3, Canada}
\altaffiliation[Present Address: ]{TRIUMF, 4004 Wesbrook Mall, Vancouver, British Columbia V6T 2A3, Canada}

\author{K.~Ortner}
\affiliation{Department of Chemistry, Simon Fraser University, Burnaby, British Columbia V5A 1S6, Canada}

\author{J.~Park}
\altaffiliation{Present Address: Department of Physics, Lund University, 22100 Lund, Sweden}
\affiliation{TRIUMF, 4004 Wesbrook Mall, Vancouver, British Columbia V6T 2A3, Canada}
\affiliation{Department of Physics and Astronomy, University of British Columbia, Vancouver, British Columbia V6T 1Z4, Canada}

\author{C.~M.~Petrache}
\affiliation{Centre de Sciences Nucl\'eaires et Sciences de la Mati\`ere, CNRS/IN2P3, Universit\'e Paris-Saclay, 91405 Orsay, France}

\author{J.~L.~Pore}
\altaffiliation{Present Address: Lawrence Berkeley National Laboratory, Berkeley, CA 94720, USA}
\affiliation{Department of Chemistry, Simon Fraser University, Burnaby, British Columbia V5A 1S6, Canada}

\author{J.~K.~Smith}
\altaffiliation{Present Address: Department of Physics, Pierce College, Puyallup, Washington, 98374, USA}
\affiliation{TRIUMF, 4004 Wesbrook Mall, Vancouver, British Columbia V6T 2A3, Canada}

\author{D.~Southall}
\altaffiliation{Present Address: Department of Physics, University of Chicago, Chicago, Illinois 60637, USA}
\affiliation{TRIUMF, 4004 Wesbrook Mall, Vancouver, British Columbia V6T 2A3, Canada}

\author{M.~Ticu}
\affiliation{Department of Chemistry, Simon Fraser University, Burnaby, British Columbia V5A 1S6, Canada}

\author{J.~Turko}
\affiliation{Department of Physics, University of Guelph, Guelph, Ontario N1G 2W1, Canada}

\author{K.~Whitmore}
\affiliation{Department of Chemistry, Simon Fraser University, Burnaby, British Columbia V5A 1S6, Canada}

\author{T.~Zidar}
\affiliation{Department of Physics, University of Guelph, Guelph, Ontario N1G 2W1, Canada}


\date{\today}

\begin{abstract}
The half-lives of three $\beta$ decaying states of $^{131}_{~49}$In$_{82}$ have been measured with the GRIFFIN $\gamma$-ray spectrometer at the TRIUMF-ISAC facility to be $T_{1/2}(1/2^-)=328(15)$~ms, $T_{1/2}(9/2^+)=265(8)$~ms, and $T_{1/2}(21/2^+)=323(55)$~ms, respectively. The first observation of $\gamma$-rays following the $\beta n$ decay of $^{131}$In into $^{130}$Sn is reported. The $\beta$-delayed neutron emission probability is determined to be $P_{1n} = 12(7)\%$ for the $21/2^+$ state and $2.3(3)\%$ from the combined $1/2^-$ and $9/2^+$ states of $^{131}_{~49}$In$_{82}$ observed in this experiment. A significant expansion of the decay scheme of $^{131}$In, including 17 new excited states and 34 new $\gamma$-ray transitions in $^{131}_{~50}$Sn$_{81}$ is also reported. This leads to large changes in the deduced $\beta$ branching ratios to some of the low-lying states of $^{131}$Sn compared to previous work with implications for the strength of the first-forbidden $\beta$ transitions in the vicinity of doubly-magic $^{132}_{~50}$Sn$_{82}$.
\end{abstract}


\maketitle


\section{Introduction}
The nuclear shell-model~\cite{mayer49,jensen49} has been of paramount importance in understanding the structure and properties of the atomic nucleus. However, as experiments have probed nuclei far from stability, it has become increasingly clear that the single-particle states of nucleons evolve as a function of isospin~\cite{otsuka01,otsuka05}. The evolution of the single-particle states in neutron-rich nuclei near traditional magic numbers is a prominent subject in current nuclear structure research~\cite{otsuka05}. A thorough understanding of the microscopic mechanisms responsible for this single-particle shell evolution is essential in order to make accurate predictions of nuclear properties far from stability.

The Sn isotopes (with magic proton number $Z=50$) provide an important testing ground for nuclear structure models as they form the longest isotopic chain accessible to experiments. The evolution of the single-particle states in the region around $^{132}_{~50}$Sn$_{82}$ is of particular interest as it is the heaviest neutron-rich doubly magic nucleus for which spectroscopic information is available~\cite{Jones2010,Jones2011}. Nuclei in the vicinity of $^{132}$Sn have thus been the subject of intense study~\cite{Jones2010,orlandi18,PhysRevLett.112.132501,PhysRevC.70.034312}. However, detailed knowledge of the nuclear structure of some isotopes remains lacking. Direct reaction experiments are a powerful tool as they are able to probe the single-particle states of nuclei in this region~\cite{kozub12,orlandi18,Jones2010,Jones2011}. Additionally, experiments such as $\beta$-decay and fission are able to probe higher energy, core-excited states in these nuclei and can provide information on correlations above and below the shell gaps~\cite{bhattacharyya01}. An expanded knowledge of nuclei in this region will provide information on the robustness of the doubly magic $^{132}_{~50}$Sn$_{82}$ core.

In addition to defining the single-neutron hole states in $^{132}$Sn, the nucleus $^{131}$Sn also plays an important role in the astrophysical rapid neutron-capture ($r$)-process~\cite{mumpower16}, which is responsible for the production of nearly half of the observed isotopes heavier than Fe~\cite{bbfh57,cameron57,cameron57-2}. In the $r$-process, the $N=82$ isotones act as waiting-points where an accumulation of $r$-process material occurs before it can be transferred to the next elemental chain via $\beta$-decay. The half-lives of these nuclei determine how much material is accumulated at each waiting point, and hence, the amplitude and shape of the resulting $r$-process abundance peaks~\cite{mumpower16,seeger65,cameron1983,cameron1983-2}. Since many $r$-process nuclei are experimentally inaccessible, calculations of the $r$-process flow rely on the predictions of nuclear physics properties such as half-lives, neutron separation energies, neutron capture cross-sections, and $\beta$-delayed neutron branching ratios of the nuclei involved in this path. The level structures of $^{131,133}$Sn have a large influence on the calculated neutron capture cross-sections in this region and thus, have important consequences for the calculated abundance of nuclei across the entire nuclear landscape~\cite{zhang12,mumpower16,Surman09}.

In neutron-rich ``cold'' $r$-process scenarios
\cite{freiburghaus99,korobkin12} such as neutron star mergers, the $r$-process reaction path is pushed towards the neutron drip-line. In this environment, the very neutron rich $N=82$ isotones below $^{132}$Sn are strongly populated and have a major influence on the $r$-process flow~\cite{mumpower16}. Many of these nuclei are beyond the reach of current experimental facilities and require theoretical models to predict the relevant $r$-process quantities, such as the $\beta$-decay half-lives. Experimental measurements of the half-lives of less exotic $N=82$ isotones are thus critical in order to benchmark the models used to predict the properties of the more neutron-rich isotopes close to the dripline. 

Recently, half-life measurements for the $N=82$ nucleus $^{130}$Cd~\cite{lorusso15,PhysRevC.93.062801} resolved a discrepancy between measurements and theoretical calculations~\cite{cuenca07,zhi13} of half-lives for the $N=82$ waiting-point nuclei $^{129}$Ag, $^{128}$Pd and $^{127}$Rh. However, in light of the reduced quenching of the Gamow-Teller (GT) transition strength implied by these recent measurements, the calculated half-life of $^{131}$In~\cite{PhysRevC.70.034312} now appears too short. The first-forbidden (FF) $\beta$-decay strength in this region used in the calculation of Ref.~\cite{zhi13} is estimated from the $\beta$ decay of the $1/2^-$ isomers of $^{129,131}$In and is believed to account for approximately 10-20\% of the decay width for the $N=82$ isotones, with the $^{131}$In decay having the largest FF contribution. It is possible that the discrepancy in the calculated half-life for $^{131}$In results because current shell-model calculations for $^{131}$Sn are unable to produce the entire spectrum of accessible $\beta$-decay levels, or because the estimated FF strength for nuclei in this region is incorrect. More detailed spectroscopic information on the levels populated in the $\beta$ decay of $^{131}$In is thus crucial to understanding the $\beta$-decay properties in the neutron-rich $N=82$ region.

\section{Experiment}
Gamma-Ray Infrastructure For Fundamental Investigations of Nuclei (GRIFFIN)~\cite{svensson14,garnsworthy17,garnsworthy18} is a high-efficiency $\gamma$-ray spectrometer comprised of 16 high-purity germanium (HPGe) clover detectors~\cite{RIZWAN2016126} located at the Isotope Separator and ACcelerator (ISAC) facility at TRIUMF~\cite{dilling13}. In the present experiment, the $^{131}$In isotope of interest was produced in the spallation of uranium, with a 9.8-$\mu$A, 480-MeV proton beam incident on a uranium carbide (UC$_x$) target. The ion-guide laser ion source (IG-LIS)~\cite{raeder14} was used to resonant laser ionize the $^{131}$In and suppress surface-ionized isobars. A low-energy (28~keV) radioactive ion beam (RIB) of $^{131}$In was selected by a high-resolution mass separator ($\delta M/M \sim 1/2000$) and delivered to the GRIFFIN spectrometer in the \mbox{ISAC-I} experimental hall. 

The one proton-hole nucleus $^{131}$In is known to have three $\beta$-decaying states with spin assignments of $1/2^-$, $9/2^+$ and $21/2^+$~\cite{PhysRevC.70.034312}. A mixture of these three states was delivered to GRIFFIN. As all three states have similar half-lives, it is difficult to assign the individual $\gamma$-ray transitions to the decay of each parent state based on time structure alone. The high photo-peak efficiency of the GRIFFIN $\gamma$-ray spectrometer, however, makes it possible to assign many of the $\gamma$-rays to a specific cascade using $\gamma$-$\gamma$ coincidences. These assignments are definitive for the $\gamma$-rays following the decay of the high-spin $21/2^+$ isomer of $^{131}$In where all of the feeding collects into excited states of $^{131}$Sn that decay via $\gamma$ rays with energies above~$4$~MeV. The $1/2^-$ and $9/2^+$ decays are, however, more challenging to assign as they populate, to varying degrees, many of the same $\gamma$-ray transitions in the $^{131}$Sn daughter nucleus.

The RIB was implanted into a mylar tape at the mutual centers of the SCintillating Electron-Positron Tagging ARray (SCEPTAR), an array of 20 thin plastic scintillators for tagging $\beta$ particles~\cite{ball05,garrett15,garnsworthy18}, and GRIFFIN~\cite{svensson14,garnsworthy17,garnsworthy18}. The mylar tape is part of a moving tape system which allows long-lived daughter nuclei and contaminant activities to be removed from the array into a lead shielded tape collection box. A single cycle for the $^{131}$In decay experiment starts by moving the tape over a duration of 1.5~s. Once the tape move was complete, a background measurement was performed for 2~s, followed by a collection period with the beam being implanted into the tape for 30~s, followed by a decay period of 5~s during which the beam was deflected by the ISAC electrostatic kicker two floors below the GRIFFIN array. Following this cycle, the tape was again moved into the lead-shielded tape collection box and the cycle was repeated. The results reported here were obtained from 182 cycles recorded over a period of approximately 1.9~hours with an average $^{131}$In beam intensity of approximately 500 ions/s for each of the $9/2^+$ and $1/2^-$ isomers, and 60~ions/s for the $21/2^+$ isomer.

Some of the $\gamma$ rays from $^{131}$In decay were contaminated by $\gamma$ rays of very similar energies following the $\beta$ decay of the two $\beta$-decaying states of $^{131}$Sn ($T_{1/2}$ = 56.0~s and 58.4~s~\cite{stone86}). The intensities of these $\gamma$-ray photopeaks from $^{131}$Sn decay were measured at the end of the experimental cycles, after the short-lived $^{131}$In activity had decayed away, specifically, in the last 2~s of the decay cycle. This region of the cycle was approximately 10 $^{131}$In half-lives after the beam was blocked and thus had an insignificant contribution of $^{131}$In decay compared to $^{131}$Sn decay. This allowed a measurement of the ratio of the areas of each $\gamma$-ray observed in this spectrum compared to the strong 798~keV $\gamma$-ray in $^{131}$Sb following $^{131}$Sn decay. The amount of contamination in the $^{131}$In photopeaks was then determined by comparing this ratio to the area of the 798~keV photopeak in the beam-on part of the spectrum. It is worth noting that the two known $\beta$-decaying states in $^{131}$Sn have approximately equal half-lives, so the relative populations of these two states at any given point in the cycle did not vary significantly. A very weak in-beam contamination of $^{131}$La ($T_{1/2} = 59(2)$~min~\cite{yaffe63,creager60}) was also observed. The level of contamination was sufficiently small that the $\gamma$ rays from this decay could only be observed when applying a $\gamma$-$\gamma$ coincidence condition. Nonetheless, care was taken to avoid assigning $\gamma$ rays from this decay to the level scheme of $^{131}$Sn.

GRIFFIN was operated in high-efficiency mode with the clovers positioned at 11~cm from the beam spot~\cite{garnsworthy18}. The data were analyzed using a clover addback algorithm in which the measured $\gamma$-ray energies in coincidence from each HPGe crystal within a single clover detector were summed. This method has the advantage of increasing both the photopeak efficiency and the peak-to-total ratio in the $\gamma$-ray spectrum by recovering the Compton-scattered $\gamma$-ray events that had full energy deposition within a single clover detector~\cite{garnsworthy18}. Unless otherwise noted, coincidences between $\gamma$ rays were constructed using a 250~ns time gate. 

The SCEPTAR array was used to tag events originating from $\beta$ decay on the mylar tape, thereby suppressing room-background $\gamma$-ray events~\cite{ball05,garrett15,garnsworthy18}. The coincidence condition between detected $\beta$ particles in SCEPTAR and $\gamma$-rays detected in GRIFFIN was set to 250~ns unless otherwise noted.

\section{Results}
\subsection{$\beta$-delayed neutron branches in $^{131}$In decay}\label{sec:b-n-decay}
\begin{figure}[!t]
	\subfigure{
   		\includegraphics[width=\linewidth]{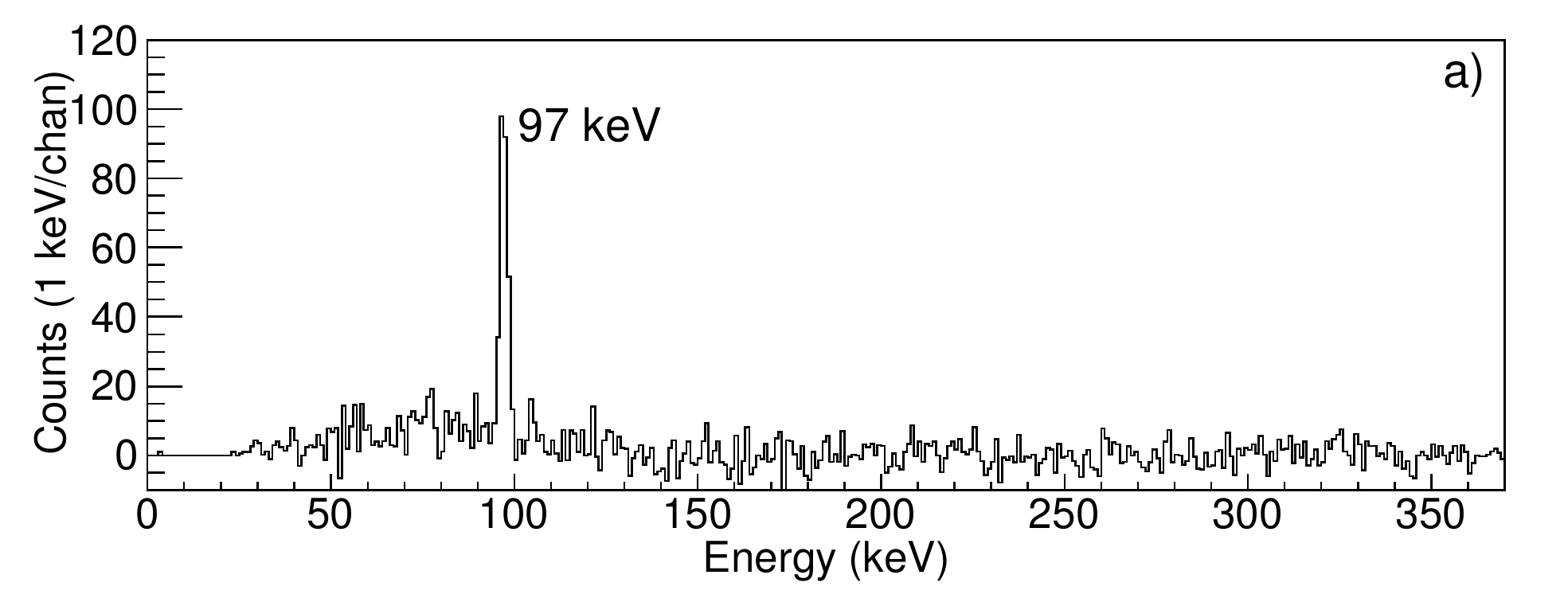}%
   		\label{fig:coinc391}
   	}
    \subfigure{
   		\includegraphics[width=\linewidth]{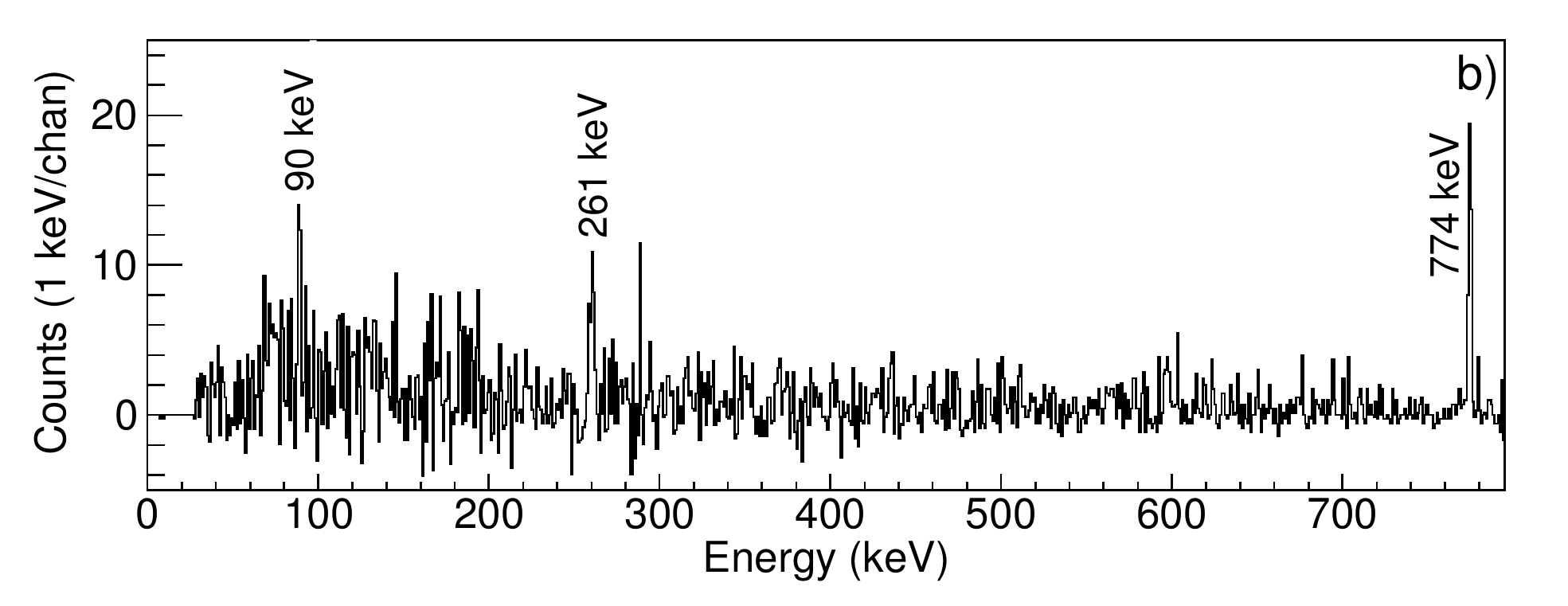}%
   		\label{fig:coinc1221}
   	}
   \caption{$\beta$-coincident $\gamma$-$\gamma$ spectra gated on a) the 391-keV and b) the 1221-keV $\gamma$ rays. These spectra show the presence of excited states of $^{130}$Sn following the $\beta n$ decay of $^{131}$In. }
   \label{fig:coinc_beta_n}
\end{figure}

\begin{figure}[t!]
   \includegraphics[width=\linewidth]{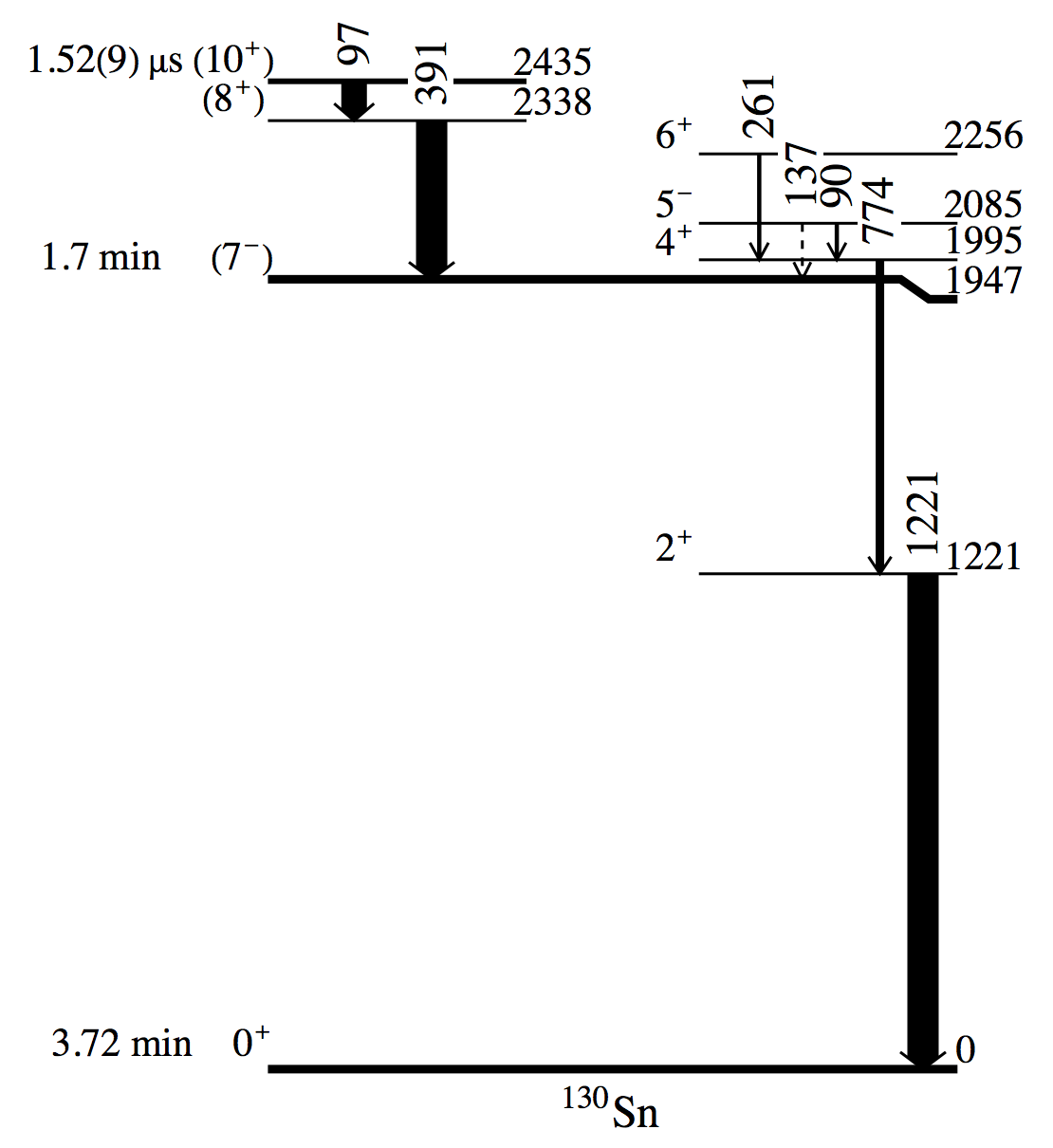}%
   \caption{The decay scheme of $^{130}$Sn observed following the $\beta n$ decay of $^{131}$In. The half-lives of the $\beta$-decaying $(7^-)$ and $0^+$ state are from Refs.~\cite{kerek74,izak72} and were used in determining the total number of $\beta n$ decays. The half-life of the $(10^+)$ isomeric state was measured in this work. See text for details.  
   \label{fig:sn130-decay-scheme}}
\end{figure}

Previous studies of the decay of $^{131}$In have measured weak $\beta n$ branches into $^{130}$Sn via neutron detection~\cite{NSR1986WA17,RUDSTAM19931}. In the current work, the $\beta n$ decay branch has been observed via $\gamma$ rays in $^{130}$Sn for the first time as shown in Fig.~\ref{fig:coinc_beta_n}. Figure~\ref{fig:sn130-decay-scheme} shows the portion of the level scheme that was observed following the $\beta n$ decay of $^{131}$In in this work. Table~\ref{tab:130Sn_intensities} shows the measured intensities of each transition in $^{130}$Sn observed in this work. 

\begin{figure}[t!]
   \includegraphics[width=\linewidth]{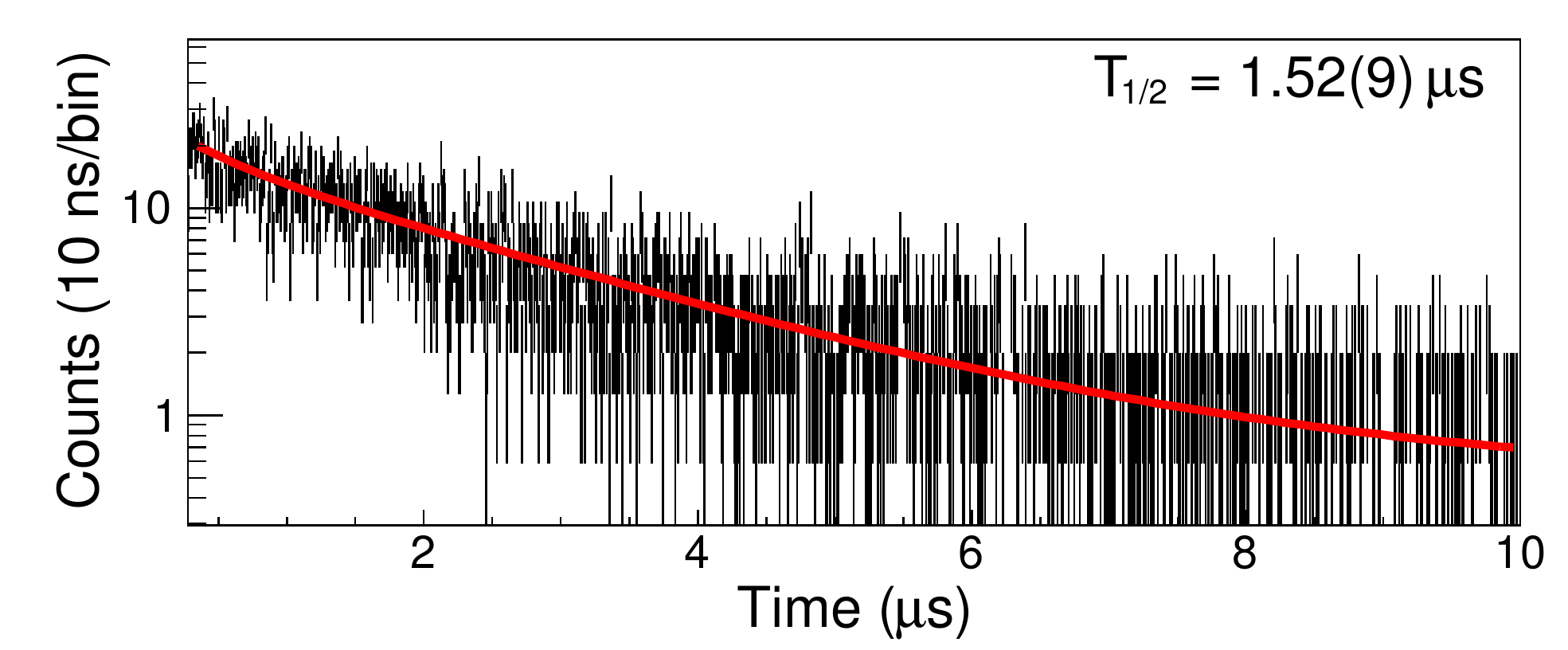}%
   \caption{(Color Online) Half-life measurement for the 2435~keV, $I^\pi = (10^+)$ isomer in $^{130}$Sn obtained from the $\beta$-$\gamma$ time difference gated on the 391-keV photopeak in $^{130}$Sn. The half-life was measured from these data to be 1.52(9)~$\mu$s in good agreement with the previous measurement of 1.61(15)~$\mu$s~\cite{FOGELBERG1981157}. This state was populated following the $\beta n$ decay of $^{131}$In. See text or details.
   \label{fig:sn130_isomer_hl}}
\end{figure}

The 2435-keV $I^\pi = (10^+)$ isomer in $^{130}$Sn was previously identified following the $\beta$-decay of $^{130}$In~\cite{FOGELBERG1981157}. The half-life of this state was measured in the current work using the $\beta$-$\gamma$ time difference between a $\beta$ particle measured in SCEPTAR and a 391-keV $\gamma$ ray measured in a GRIFFIN detector and is shown in Fig.~\ref{fig:sn130_isomer_hl}. The fit included a constant background as well as a Compton scattered $\gamma$-ray background component with the 316(5)~ns half-life of the high-spin isomer in $^{131}$Sn measured in this work. This fit yielded a half-life of 1.52(9)~$\mu$s, in good agreement with, but more precise than the previous measurement of 1.61(15)~$\mu$s~\cite{FOGELBERG1981157}. A ``chop analysis'' was performed where the fit was performed over subsets of the total range in order to investigate any potential systematic effects of the fit range including any rate-dependent effects such as dead time, and none were found. The half-life of the $^{131}$Sn background component was also varied from 280~ns to 340~ns, with no significant effect on the measured half-life of the 2435-keV isomer (1.49--1.53~$\mu$s).  

The $\gamma$-ray intensities from the decay of the 2435~keV state were obtained from $\beta$-$\gamma$ coincidence photopeak areas and were corrected for the missing events that lie outside of the coincidence time window. This was done for the entire intensity of the 97-keV $\gamma$ ray (with a theoretical internal conversion coefficient of $\alpha = 1.4(4)$~\cite{kibedi08}), while only the component of the 391-keV intensity that was fed by the 97-keV transition was corrected.
\begin{table}[t!]
\begin{center}
    \caption{The $\gamma$ rays observed in $^{130}$Sn following the $\beta n$ decay of $^{131}$In. $N_\gamma$ represents the number of decays detected through this transition, corrected for the $\gamma$-ray efficiency. This also includes the internal conversion coefficient, and a correction for the missed $\beta$-$\gamma$ coincidences due to the isomeric state at 2435~keV.} 
    \sisetup{table-align-text-post = false}
    \begin{tabular}{  l  S[table-format=5.5]  S[table-format=2.7] S[table-format=2.6] }
    \hline
    \multicolumn{1}{c}{\multirow{3}{*}{\shortstack[c]{Level\\Energy\\(keV)}}} & \multicolumn{1}{c}{\multirow{3}{*}{\shortstack[c]   {$\gamma$-ray\\Energy\\(keV)}}} & &  \\
    
  &  &  \multicolumn{1}{c}{\multirow{2}{*}{\shortstack[c]{$N_\gamma$\\($\times10^4$)}}} & 
  
  \multicolumn{1}{c}{\multirow{2}{*}{\shortstack[c]{$N_{\gamma,_{out}} - N_{\gamma,_{in}}$\\($\times10^4$)}}} \\	
   
   & & &					\\
    \hline\hline
    1221.13(20) 		& 1221.13(20) 	& 2.58(11) & 2.38(12)\\
    1995.39(25)  		& 774.26(24)	& 0.20(4)  & 0.06(6) \\
    2085.5(4)   		& 137.96(5)\textsuperscript{\emph{b}}			& 0.042(17)\textsuperscript{\emph{a}} & 0.12(4) \\ 
        			    & 90.2(4)		& 0.074(31) & \\
	2256.45(28) 		& 261.1(3)		& 0.067(27)& 0.067(27)\\
    2338.27\textsuperscript{\emph{b}}	& 391.38(20)& 2.16(7) & 0.61(22)	\\
    2434.81\textsuperscript{\emph{b}}	& 97.24(25)	& 1.55(21) & 1.55(21) \\\hline\hline
    \end{tabular}
    \begin{flushleft}
    \textsuperscript{\emph{a}} Deduced from the intensity of the 90-keV $\gamma$-ray and the branching ratio of Ref.~\cite{FOGELBERG1981157}.\\
    \textsuperscript{\emph{b}} Energy from Ref.~\cite{FOGELBERG1981157}.\\
    \end{flushleft}
    \label{tab:130Sn_intensities}
\end{center}
\end{table}

\begin{figure}[t!]
   \includegraphics[width=\linewidth]{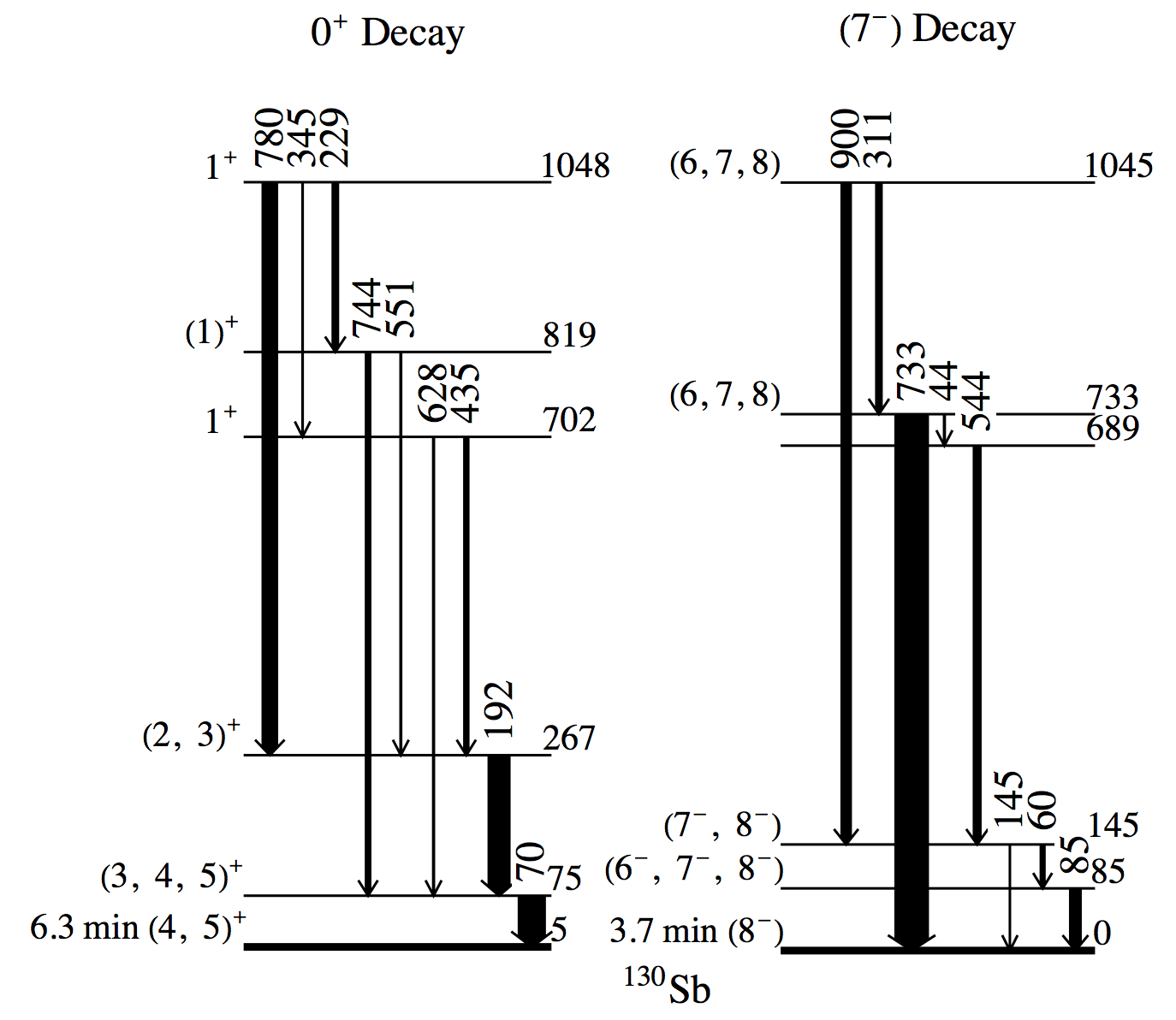}%
   \caption{A partial decay scheme of $^{130}$Sb following the $\beta$ decay of the $0^+$ and $(7^-)$ states of $^{130}$Sn. The intensities of these $\gamma$ rays were used to determine the total number of $\beta$ decays of $^{130}$Sn observed in this work, which is related to the total number of $\beta n$ decays of $^{131}$In. 
   \label{fig:Sb130Scheme}}
\end{figure}

The previously evaluated value for the $\beta$-delayed neutron emission probability, $P_{1n}$, for the $21/2^+$ decay of $^{131}$In was 0.028(5)\%~\cite{ENSDFIn131}, while the $P_{1n}$ value for the mixed decay of the $1/2^-$ and $9/2^+$ decays is quoted as 2.0(4)\%~\cite{NSR1986WA17,RUDSTAM19931,ENSDFIn131}. To constrain the total number of $\beta n$ decays of $^{131}$In observed in this experiment, $\gamma$ rays in $^{130}$Sb following the $\beta$ decay of $^{130}$Sn were used. The $\beta n$ decay of $^{131}$In will, after it deexcites, eventually populate one of either the $0^+$ ground state ($T_{1/2} = 3.72$~min) or the $(7^-)$ 1947-keV isomer ($T_{1/2} = 1.7$~min) of $^{130}$Sn. The $\gamma$-ray intensities following the $\beta$ decays of the $(7^-)$ isomeric and $0^+$ ground state of $^{130}$Sn are well known~\cite{NSR1994WAZU} and were used to determine the number of decays from each of these states. Due to the long half-lives of the $\beta$-decaying states of $^{130}$Sn compared to the cycle time, not all of the $^{130}$Sn created by $\beta n$ decay would have decayed over the course of the cycle. Therefore the measured intensities were corrected for the implantation and collection times of the cycle with respect to the half-life of the particular state of origin for these $\gamma$ rays. The $\gamma$ rays used in this analysis are shown in a partial level scheme in Fig.~\ref{fig:Sb130Scheme}. From the intensity of the 733-keV $\gamma$ ray in $^{130}$Sb, the number of $\beta$ decays from the 1947-keV $(7^-)$ isomeric state of $^{130}$Sn was measured to be $6.2(20)\times10^4$. As a check, the intensities of the coincidences between the 145-keV and 544-keV $\gamma$ rays and the coincidences between the 311-keV and 733-keV $\gamma$ rays in $^{130}$Sb were also used. For each coincidence mentioned, the $\gamma$-ray branching ratios and internal conversion were considered where necessary. The number of $^{130}$Sn decays measured using these coincidences were $6.6(23)\times10^4$ and $6.2(22)\times10^4$, respectively, in good agreement with each other and with the value of $6.2(20)\times10^4$ deduced from the intensity of the 733-keV $\gamma$ ray.

The total number of $^{130}$Sn $0^+$ ground state decays was measured using a combination of the 435-192 and 192-70~keV coincidences resulting in $6.4(15)\times10^4$ and $5.9(13)\times10^4$ $\beta$ decays, respectively. The number of $\beta$ decays from this state was determined from the weighted average to be $6.1(8)\times10^4$. Therefore, the total number of $\beta$ decays from the $0^+$ and $(7^-)$ states of $^{130}$Sn measured during this experiment was $1.25(25)\times10^5$.

The ability of GRIFFIN to detect the $\gamma$ rays in $^{130}$Sn following the $\beta n$ decay of $^{131}$In provides excellent sensitivity to the isomeric origin of the $\beta n$ feeding. For example, following a $\beta$-decay from the $9/2^+$ state of $^{131}$In, in order to populate the $(10^+)$ state in $^{130}$Sn, a total orbital angular momentum of $l=4\hbar$ would have to be carried away by the neutron, $\beta^-$ particle and neutrino. However, the much more probable scenario is that this state is populated following the $\beta n$ decay of the $21/2^+$ state in $^{131}$In with no orbital angular momentum transferred to the emitted particles. Therefore, the minimum number of $\beta n$ events from the $21/2^+$ state in this work was determined from the intensity of the 97-keV $\gamma$ ray ($I_{97}$) following the $\gamma$ decay of the $(10^+)$ state in $^{130}$Sn to be $\ge1.55(21)\times10^4$. Comparing this to the number of $21/2^+$ $\beta$ decays that do not emit neutrons, $I_\beta(21/2^+) = 3.59(12)\times10^5$, as described in Sec.~\ref{sec:high-spin}, leads to a firm lower limit of $P_{1n}(21/2^+) \ge I_{97}/(I_{97} + I_\beta) = 4.2(6)\%$, in stark contrast with the value of $P_{1n}(21/2^+) = 0.028(5)\%$ reported in Ref.~\cite{ENSDFIn131}. This previous $P_{1n}$ value was estimated from fission yields in the evaluation of Ref.~\cite{RUDSTAM19931}, a method that in the meantime is seen as providing quite unreliable results.

The $\beta n$ decay to the $(8^+)$ state at 2338~keV in $^{130}$Sn from both the $9/2^+$ and $21/2^+$ states in $^{131}$In requires an orbital angular momentum transfer of $l\ge2\hbar$. However, with {$Q(\beta n)=4037(5)$~keV}~\cite{PhysRevLett.109.032501} for the decay of the $9/2^+$ ground state of $^{131}$In to $^{130}$Sn, populating the $(8^+)$ state at 2338~keV leads to a maximum neutron energy of 1.7~MeV in the case of zero kinetic energy carried away by the $\beta^-$ particle and neutrino. It is highly improbable for such low-energy neutrons to carry enough angular momentum to populate the high-spin $(8^+)$ state. Therefore, the observed population of this state is likely to be dominated by the $\beta n$ decay of the $21/2^+$ state in $^{131}$In which has a $Q_{\beta n}$ value 3764(88)~keV higher than the $9/2^+$ ground state. The number of $\beta n$ events from the $21/2^+$ state in this work was thus estimated from the intensity ($I_{391}$) collected into the 391-keV $\gamma$ ray following the $\gamma$ decay of the $(8^+)$ and $(10^+)$ states in $^{130}$Sn to be $\ge2.16(7)\times10^4$, or $P_{1n}(21/2^+)\ge I_{391}/(I_{391} + I_\beta) = 5.7(3)\%$. Considering that the rest of the intensity from the $\beta n$ decay of the $21/2^+$ state in $^{131}$In likely collects into the $(7^-)$ state in $^{130}$Sn, an upper limit on the number of $\beta n$ decays from this state into $^{130}$Sn is $P_{1n}(21/2^+)\le15(4)\%$. This is an upper limit due to the unknown population of the $(7^-)$ state from the $\beta n$ decay of the $9/2^+$ state in $^{131}$In. For the purposes of reporting $\beta$-decay branching ratios and $\gamma$-ray intensities, we therefore adopt the number of $\beta n$ decays from the $21/2^+$ state to be $5.1(31)\times10^4$, or $P_{1n}(21/2^+) = 12(7)\%$ which encompasses both limits. 

A lower limit on the $P_{1n}$ value from the decay of the mixed $9/2^+$ and $1/2^-$ populations can be determined from the remaining observed $\gamma$ decay collected into the long-lived states of $^{130}$Sn. The previously reported weak 137-keV $\gamma$-ray transition from the $I^\pi = 5^-$ state at 2085~keV excitation energy in $^{130}$Sn to the 1947-keV $(7^-)$ isomeric state was not observed in the current experiment. This transition is, however, important for setting a lower limit on the $\beta n$ branch into $^{130}$Sn as the collected intensity terminates at the 1947-keV $\beta$-decaying state. We thus use the $\gamma$-ray branching ratios from the 2085-keV state measured in Ref.~\cite{FOGELBERG1981157}, and infer the strength of the 137-keV transition from the measured intensity of the 90-keV transition from this state. The 1946.88(10) keV excitation energy of the $(7^-)$ isomer is also taken from Ref.~\cite{FOGELBERG1981157}. Using the $\gamma$-ray intensities of the 137- and 1221-keV transitions we are thus able to measure a lower limit on the $\beta n$ decays from the mixed population to be $2.63(12)\times10^4$. Together with the combined number of $\beta$ decays from Sec.~\ref{sec:number_b_decays} measured in this work, this gives a mixed $P_{1n} > 0.532(24)\%$ for the $9/2^+$ and $1/2^-$ states. This is a lower limit on the $P_{1n}$ value as it does not account for any direct $\beta n$ feeding into the long-lived ($7^-$) isomeric state or $0^+$ ground state of $^{130}$Sn, or any $\gamma$-ray transitions into these same states from highly-excited weakly-populated states~\cite{Hardy77} that were not observed in the current experiment. Subtracting the detected $21/2^+$ $\beta n$ decays from the total detected $\beta n$ decays gives an upper limit for the number of detected $\beta n$ decays from the mixed population of $<1.10(25)\times10^5$ and a $P_{1n}<2.3(3)\%$ in good agreement with the previously evaluated value of $P_{1n}=2.0(4)\%$~\cite{NSR1986WA17,RUDSTAM19931,ENSDFIn131}. Based on the roughly equal populations of the $9/2^+$ and $1/2^-$ isomers measured in this work, we set firm upper limits on the $P_{1n}(9/2^+) < 4\%$ and $P_{1n}(1/2^-) < 4\%$, while the conservative estimate of $P_{1n} = 2(2)\%$ will be used for determining $\beta$ branching ratios of these states.

\subsection{Decay scheme of the high-spin $21/2^+$ isomer of $^{131}$In}\label{sec:high-spin}
\begin{figure*}[t!]
   \includegraphics[width=\linewidth]{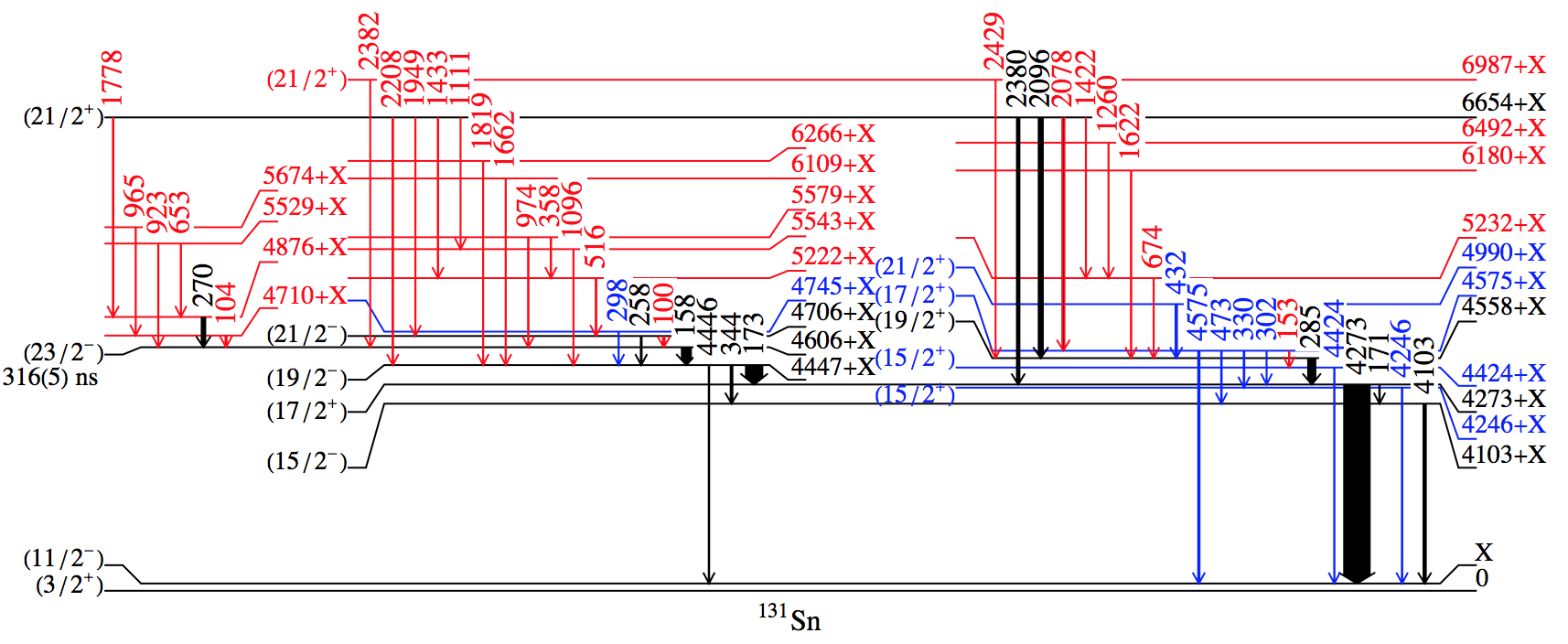}%
   \caption{(Color Online) The decay scheme of $^{131}$Sn observed following the $\beta$ decay of the $21/2^+$ isomeric state in $^{131}$In. The widths of the arrows represents the relative intensity of each transition. The low-lying isomeric state is denoted with the energy $X$ although evidence is presented in the current work that this state has an excitation energy of $X = 65.1(5)$~keV. Previously unobserved transitions and excited states are highlighted in red, while levels and $\gamma$-ray transitions which were previously identified in fission decays~\cite{bhattacharyya01} are highlighted in blue.
   \label{fig:sn131-high-spin}}
\end{figure*}

The level scheme of $^{131}$Sn populated following the $\beta$ decay of the $21/2^+$ state in $^{131}$In is shown in Fig.~\ref{fig:sn131-high-spin} and features 13 newly observed excited states and 24 newly observed $\gamma$-ray transitions. Furthermore, 5 of these levels and 8 of these $\gamma$-ray transitions which were previously identified in fission decays~\cite{bhattacharyya01} were observed following $\beta$ decay for the first time.

\begin{figure}[t!]
   \includegraphics[width=\linewidth]{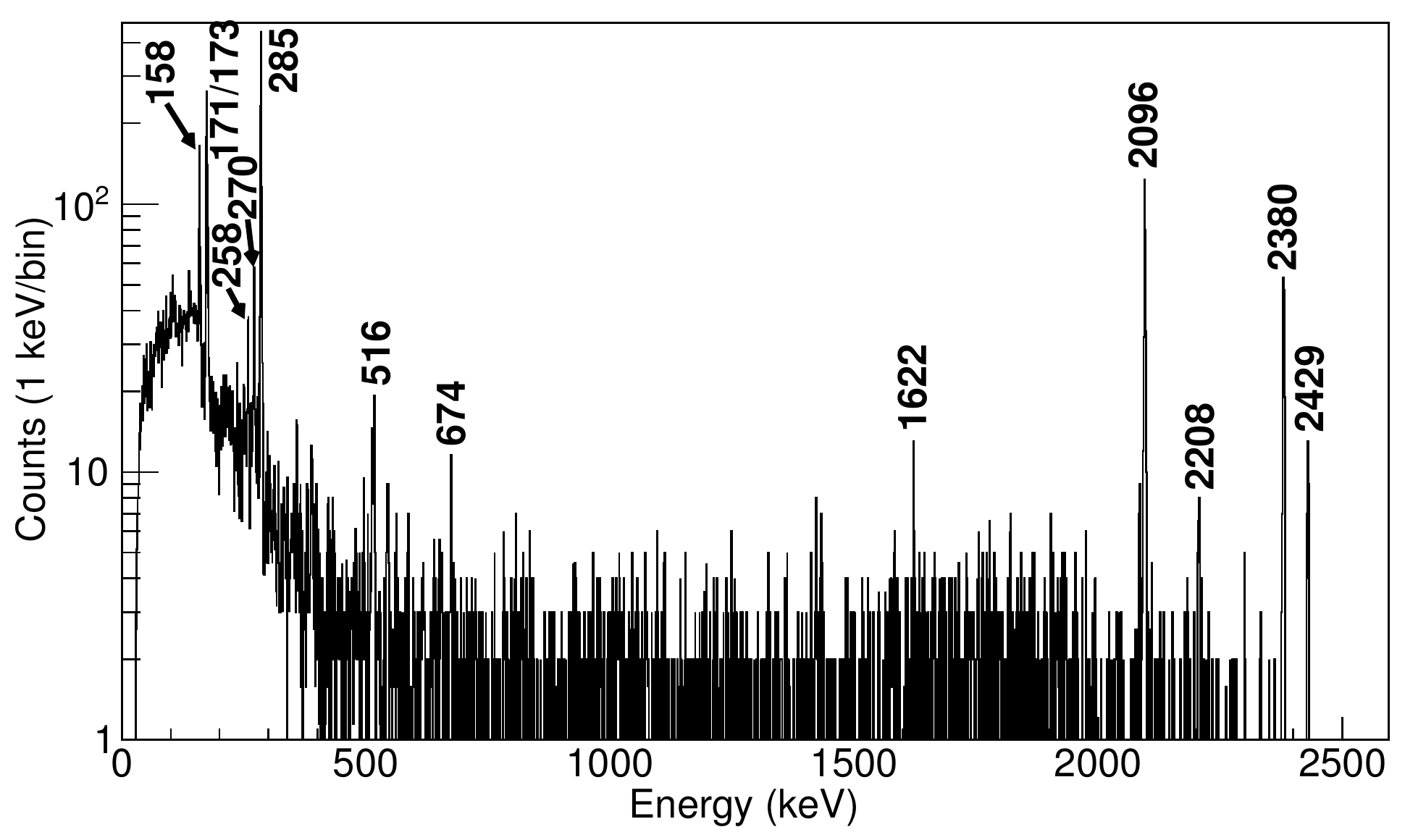}%
   \caption{The $\beta$-$\gamma$-$\gamma$ coincidence spectrum gated on the 4273~keV $\gamma$-ray transition following the $\beta$ decay of the $21/2^+$ isomer in $^{131}$In. The most intense transitions in the spectrum from $^{131}$Sn are labeled with their measured energies.
   \label{fig:4273coinc}}
\end{figure}

The $\gamma$-ray transitions in $^{131}$Sn following the $\beta$ decay of the high-spin $21/2^+$ isomer in $^{131}$In were assigned using $\gamma$-$\gamma$ coincidences with transitions that collect into the yrast high-spin states populated following this $\beta$ decay. Figure~\ref{fig:4273coinc} shows an example of the detected coincidences with the 4273~keV $\gamma$-ray transition.

It should be noted that in this work the candidate $M2$ transition of 2369~keV that was proposed~\cite{PhysRevC.70.034312} to decay from the 2434~keV $7/2^+$ excited state to the $\approx 65$~keV $\nu h^{-1}_{11/2}$ hole state in $^{131}$Sn (see Fig.~\ref{fig:sn131-med-spin} below) was observed, as shown in Fig.~\ref{fig:2369-gamma}. 
\begin{figure}[t!]
   \includegraphics[width=\linewidth]{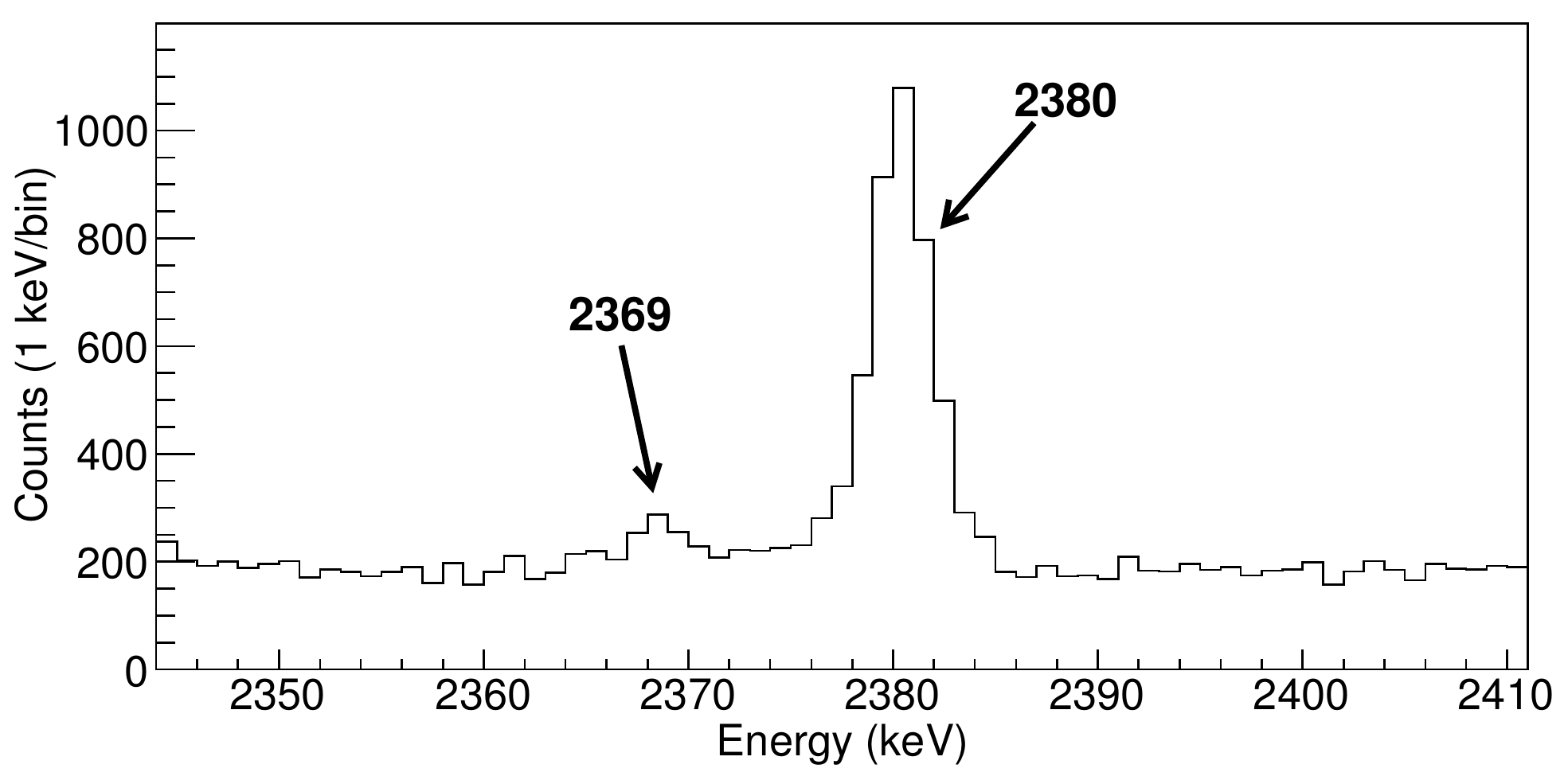}%
   \caption{The $\beta$-$\gamma$ addback spectrum showing the candidate 2369~keV $M2$ transition in $^{131}$Sn. This $\gamma$-ray is proposed to decay from the 2434~keV $\nu g_{7/2}^{-1}$ state to the 65~keV $\nu h_{11/2}^{-1}$ state. A future confirmation of the placement of this transition via $\gamma$-$\gamma$ coincidences would result in the firm placement of the $\nu h_{11/2}$ single-particle state at 65.1(5)~keV.
   \label{fig:2369-gamma}}
\end{figure}
However, due to the very weak $\gamma$ rays feeding into the 2434~keV level from above, using only the $\gamma$ rays observed in this work, it would require roughly 50-100 times more statistics to definitively confirm the placement of this $\gamma$ ray using $\gamma$-$\gamma$ coincidences. Accepting the placement of the 2369~keV $\gamma$-ray transition proposed in Ref.~\cite{PhysRevC.70.034312}, the energy of the $11/2^-$ neutron hole state was determined to be 65.1(5)~keV. As this assignment remains tentative, however, this level is denoted with an energy of $X$ for the remainder of this work.

\begin{figure}[t!]
   \includegraphics[width=\linewidth]{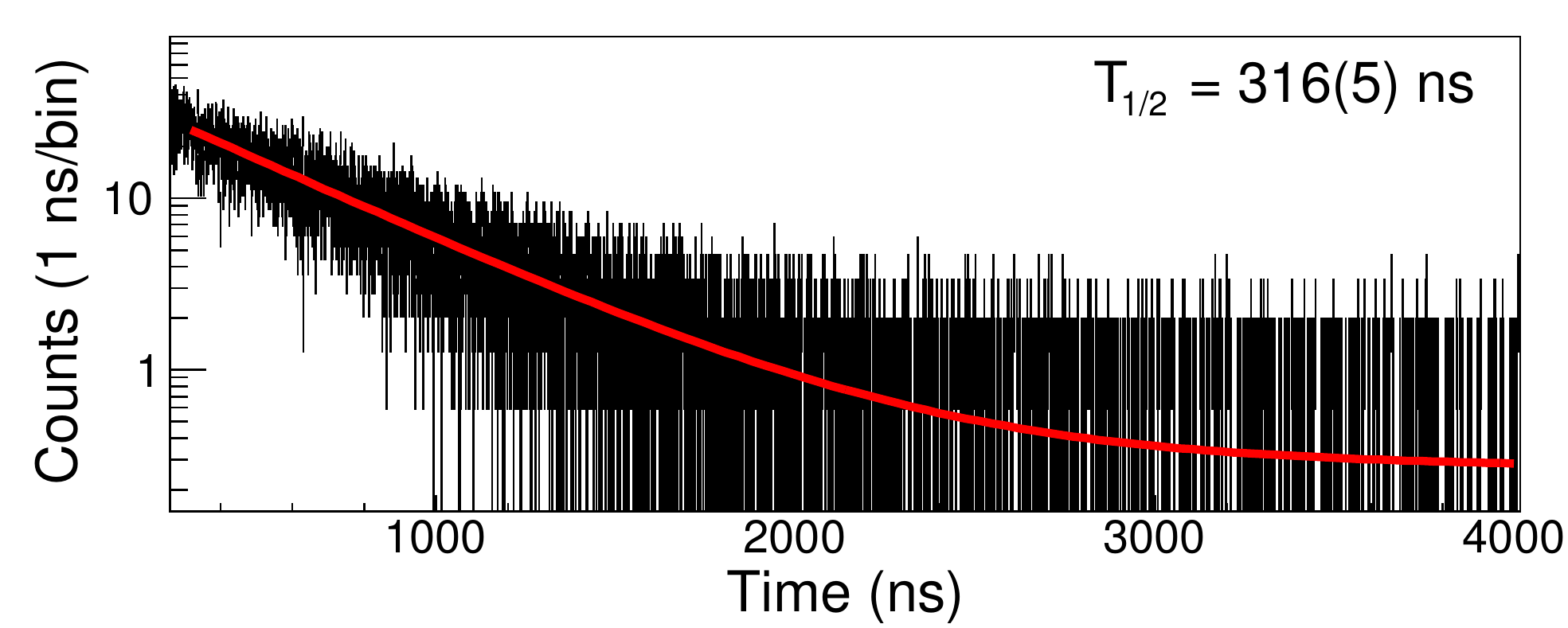}%
   \caption{(Color online) Fit of the $\beta$-$\gamma$ time difference spectrum gated on the 158-keV $\gamma$ ray. The half-life of the $4606+X$~keV high-spin isomer measured from this fit is 316(5)~ns. The fit included a component for the 1.5~$\mu$s isomer in $^{130}$Sn resulting from Compton-scattered $\gamma$ rays as well as a constant background.
   \label{fig:isomer_fit}}
\end{figure}

The half-life of the $4606+X$~keV high-spin isomer populated in $^{131}$Sn~\cite{PhysRevC.70.034312,bhattacharyya01} was measured using the time difference between a $\beta$ particle detected in SCEPTAR, and the 158-keV $\gamma$ ray depopulating this isomeric state measured in GRIFFIN. The half-life of this isomeric state was determined using data in the region from 300~ns to 10~$\mu$s following the $\beta$ trigger to ensure that the prompt-timing coincidence data were not included in the fit region. The fit included a constant background, as well as a 1.52~$\mu$s half-life component from Compton scattered $\gamma$-rays from the decay of the 2435-keV isomer in $^{130}$Sn. As shown in Fig.~\ref{fig:isomer_fit}, the half-life of the high-spin isomer in $^{131}$Sn was measured to be 316(5)~ns, in good agreement with but a factor of 4 more precise than the value of 300(20)~ns reported in Ref.~\cite{mach01}. 

The population of the $4606+X$~keV, $23/2^-$ high-spin isomer in $^{131}$Sn, either through direct feeding by $\beta$-decay or $\gamma$-ray feeding from above, presents a challenge for constructing coincidences involving this state because its half-life is roughly equal to the length of the coincidence gate used in the majority of this analysis. In order to reconstruct the coincidences across this state, a large coincidence window of 1.5~$\mu$s was used. The time-random background in this gate was removed by appropriately scaling and subtracting the number of random coincidence events that occurred between a time of 3--8~$\mu$s following the start of an event. 

\begin{figure}[!t]
   \includegraphics[width=\linewidth]{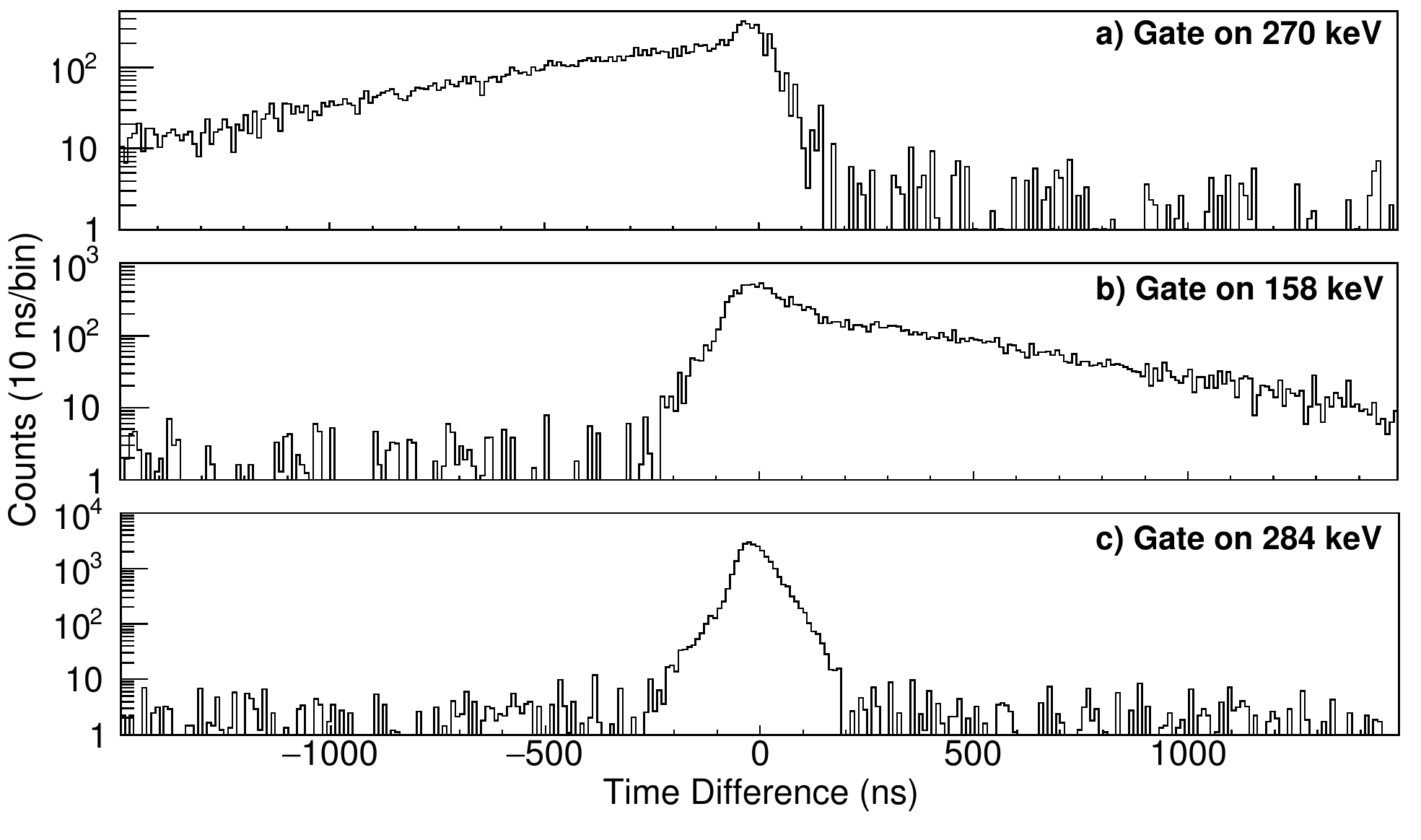}%
   \caption{The difference between the time that any $\gamma$ ray was detected relative to the a) 270-keV, b) 158-keV and c) 285-keV $\gamma$ rays. This method was used in conjunction with the $\gamma$-ray energy differences and coincidence relationships to assign the locations in the level scheme of the $\gamma$-transitions and excited states of $^{131}$Sn populated following the decay of the $21/2^+$ isomer of $^{131}$In. See text for details.
   \label{fig:isomer_timing}}
\end{figure}

The presence of the high-spin isomer, however, also offered a unique advantage in that it allowed for the analysis of the time distribution of $\gamma$-$\gamma$ and $\beta$-$\gamma$ coincidences in order to discern the time-ordering of the $\gamma$-ray transitions participating in a cascade that involved the $23/2^-$ isomeric state. For example, Fig.~\ref{fig:isomer_timing} shows the $\gamma$-$\gamma$ time difference between any $\gamma$ ray that was detected in the GRIFFIN $\gamma$-ray spectrometer with: a) a $\gamma$-ray transition feeding the $23/2^-$ isomeric state from above (270~keV), b) a $\gamma$ ray decaying from the isomeric state below (158~keV), and c) a $\gamma$ ray (285~keV) from a cascade independent of the $4606+X$~keV state. A negative time-difference ``tail'' occurs for $\gamma$-rays that are delayed relative to the gating $\gamma$ ray. This is an indication that the gating $\gamma$-ray transition takes part in a cascade feeding the isomer from above. A positive time-difference ``tail'' is evidence that the gating $\gamma$-ray transition is involved in a cascade below the isomer. This technique significantly increased the experimental sensitivity to the many low energy $\gamma$ rays feeding the isomer from above.

The multipolarities of many of the low-energy transitions in the high-spin portion of the decay scheme are uncertain. In order to determine the total intensities of these transitions, an estimate of the conversion coefficients is required. The conversion coefficients for these transitions were calculated using the frozen-orbital approximation~\cite{kibedi08} assuming each of $E1$, $M1$, and $E2$ multipolarities. In order to take into account the range of potential conversion coefficients, the adopted value in this work used the average of the $E1$ and $E2$ conversion coefficients, with an uncertainty assigned that encompassed the value from both $E1$ and $E2$. Table~\ref{tab:conversion_coeffs} shows the conversion coefficients that were adopted in the analysis of the low energy transitions. The conversion coefficient of the 100-keV transition connecting the $4706+X$- and $4606+X$-keV levels, in particular, was measured from $\gamma$-ray coincidences with the 516-keV $\gamma$ ray which feeds the $4706+X$-keV level directly from above. The intensity from the 516-keV transition must therefore pass through the 100-keV and 258-keV transitions via $\gamma$ rays or conversion electrons, with the conversion coefficient of the 258-keV $\gamma$ ray being very small. By comparing the intensity of the 100-keV and 258-keV $\gamma$ rays in coincidence with the 516-keV transition with the intensity of the 516~keV in the singles spectrum, and corrected for $\gamma$-ray detection efficiency, the ``missing'' 100-keV transition can be measured. The intensity that is not in coincidence with the 100- or 258-keV transitions is due to conversion. The measured conversion coefficient for this transition $\alpha = 0.5(5)$ suggests mainly $M1$ or $E1$ character ($\alpha(E1) = 0.180(3), \alpha(M1) = 0.557(9), \alpha(E2) = 1.58(3)$). 
\begin{table}[!t]
\begin{center}
    \caption{ Adopted conversion coefficients used in analyzing the level scheme following the $\beta$ decay of the $21/2^+$ state of $^{131}$In. The conversion coefficients were calculated using the BrICC frozen orbital approximation~\cite{kibedi08}. Due to lack of knowledge of the transition multipolarities, the conversion coefficients were estimated from the average of the $E1$ and $E2$ values and assigned an uncertainty that encompassed both the $E1$ and $E2$ possibilities unless otherwise stated.} 
    \sisetup{table-align-text-post = false}
    \begin{tabular}{  l  S[table-format=<2.7,table-align-comparator = false] S[table-format=<2.7,table-align-comparator = false]   S[table-format=<2.7,table-align-comparator = false]  }
        \hline
  \multicolumn{1}{c}{\multirow{2}{*}{\shortstack[c]{Energy\\(keV)}}} & \multicolumn{1}{c}{\multirow{2}{*}{\shortstack[c]{$\alpha_e$\\$(E1)$ }}} & \multicolumn{1}{c}{\multirow{2}{*}{\shortstack[c]{$\alpha_e$\\$(E2)$ }}} & \multicolumn{1}{c}{\multirow{2}{*}{\shortstack[c]{Adopted\\$\alpha_e$ }}} \\
      &\\
    \hline\hline
    100.35(29)		&0.180(3)       &1.576(22)   &\le0.5(5)\textsuperscript{\emph{a}}\\
    103.88(24)		&0.1629(23)     &1.394(20)  &0.78(61)	\\
    152.9(4)		&0.0546(8)      &0.355(5)   &0.20(15)	\\
    158.49(20)		&0.0493(7)      &0.313(5)   &0.18(13)  	\\
    170.92(21)		&0.0399(6)      &0.240(4)   &0.14(10)	\\
    173.26(20)		&0.0384(6)      &0.229(4)   &0.13(10)   		\\
	258.16(21)		&0.01279(18)    &0.0583(9)  &0.035(23)			\\
    270.25(20)		&0.01131(16)    &0.0500(7)  &0.031(19)			\\
	284.58(20)		&0.00985(14)    &0.0422(6)  &0.026(16)			\\
	297.73(22)		&0.00874(13)    &0.0363(5)  &0.023(14)			\\
	302.0(5)		&0.00842(12)    &0.0347(5)  &0.021(13)			\\
    329.50(22)		&0.00670(10)    &0.0262(4)  &0.017(10)			\\
    344.27(20)		&0.00599(9)     &0.0228	(4) &0.014(8)			\\ 
    \hline\hline
    \end{tabular}
    \\
    \begin{flushleft}
    \textsuperscript{\emph{a}} Measured from coincidence with the 516-keV transition.\\
    \end{flushleft}
    \label{tab:conversion_coeffs}
\end{center}
\end{table}

In Ref.~\cite{PhysRevC.70.034312}, the 4246- and 4575-keV $\gamma$ rays were placed within the decay scheme of $^{131}$Sn populated by the decay of the $1/2^-$ state of $^{131}$In due to an apparent coincidence of the 4246- and 332-keV transitions. However, in this work, it was observed that the coincidence is actually between the 4246-keV $\gamma$ ray and a 330-keV $\gamma$ ray which was first noted in Ref.~\cite{bhattacharyya01}. Furthermore, both of these $\gamma$ rays are in coincidence with the 2078-keV $\gamma$ ray, resulting in a firm reassignment of these $\gamma$ rays to the high-spin portion of the $^{131}$Sn level scheme following the $\beta$ decay of the $21/2^+$ state of $^{131}$In, as shown in Fig.~\ref{fig:sn131-high-spin}.

The 258-keV $\gamma$ ray was also observed to be in coincidence with the 173-keV and 344-keV $\gamma$ rays, but there was no observation of the 408-keV $\gamma$ ray proposed in Ref.~\cite{bhattacharyya01}. The 258-keV $\gamma$ ray was also observed to be in strong coincidence with a 1949-keV $\gamma$ ray originating from the previously established $6654+X$-keV level. Furthermore, a 100-keV transition connecting the $4606+X$- and $4706+X$-keV levels was observed to be in coincidence with the 1949- and 158-keV $\gamma$ rays, firmly establishing the position of the 258-keV $\gamma$ ray observed in this work. It is possible that another $\gamma$ ray with an energy very similar to 258~keV that is not populated in the $\beta$ decay of $^{131}$In explains the decay pattern reported in Ref.~\cite{bhattacharyya01}.

Every $\beta$ decay from the $21/2^+$ state in $^{131}$In that is not followed by neutron emission to $^{130}$Sn, is followed by $\gamma$-ray transitions in $^{131}$Sn as the direct $\beta$ decay to the low-lying $\beta$-decaying states of $^{131}$Sn is strongly forbidden. Each $\gamma$-ray cascade from this high-spin portion of the level scheme was observed to eventually feed the $11/2^-$ isomeric state. Therefore, $\beta$ feeding into the high-spin portion of the level scheme can be determined by summing the total $\gamma$-ray intensity collected into the $11/2^-$ isomeric state. This was measured to be $3.59(12)\times10^5$ events. Adding the potential number of $\beta n$ decays from this state of $5.1(31)\times10^4$ deduced in Sec.~\ref{sec:b-n-decay}, the total number of $\beta$ decays from $21/2^+$ state was determined to be $4.10(33)\times10^5$ compared to the $5.308(18)\times10^6$ total $^{131}$In $\beta$-decay events detected in this experiment. The intensities of each of the $\gamma$ rays that were observed following the decay of the $21/2^+$ isomer are shown in Table~\ref{tab:high_spin_intensities}.
\begin{table}[t!]
\begin{center}
    \caption{The observed excited states and $\gamma$-ray transitions in $^{131}$Sn following the $\beta$-decay of the $21/2^+$ isomer in $^{131}$In. For the $\gamma$ decaying states, $I_{out}-I_{in}$ represents the excess observed $\gamma$ ray and conversion electron intensity out of a state compared to the intensity into the state. This represents an upper limit for the $\beta$ feeding. } 
    \sisetup{table-align-text-post = false}
    \begin{tabular}{  l  S[table-format=<3.5,table-align-comparator = false]  S[table-format=5.6]  S[table-format=3.6] }
    \hline
    \multicolumn{1}{c}{\multirow{3}{*}{\shortstack[c]{Level\\Energy$+X$\\(keV)}}}	& & \multicolumn{1}{c}{\multirow{3}{*}{\shortstack[c]   {$\gamma$-ray\\Energy\\(keV)}}}  & \multicolumn{1}{c}{\multirow{3}{*}{\shortstack[c]{Absolute\\Intensity\\(\%)$^c$}}}  \\ 
    
    & \multicolumn{1}{c}{\multirow{2}{*}{\shortstack[c]{$I_{out}$ - $I_{in}$\\(\%)$^c$}}} &  &   \\	
    
   & & &					\\
    \hline\hline
    4102.75(26) & \le 1.0		& 4102.75(26)\textsuperscript{\emph{b}}	& 6.7(12)\textsuperscript{\emph{b}}\\
    4245.76(25) & \le0.17 		& 4245.76(25) 	& 1.64(19)\\
    4273.20(20) & \le8.8		& 170.92(21)\textsuperscript{\emph{b}}  	& 1.27(19)\textsuperscript{\emph{b}}\\
    			& 				& 4273.20(20) 	& 73(3)\\
    4424.2(6) 	& \le0.22		& 4424.2(6)   	& 0.66(18)\\       
    4446.9(4) 	& 9(4)			& 173.26(20)\textsuperscript{\emph{b}}  	& 32(3)\\         
				& 				& 344.27(20)  	& 4.52(21)\\
                & 				& 4447.2(5)   	& 1.3(8)\\
	4557.8(4) 	& 3.3(8)		& 284.58(20)  	& 28.0(10)\textsuperscript{\emph{a}}\\
	4575.49(24) & 2.1(5)		& 152.43(4)	   	& 0.62(13)\\
    			& 				& 302.0(5)	   	& 0.30(16)\textsuperscript{\emph{a}}\\
                & 				& 329.50(22)\textsuperscript{\emph{b}}	& 1.70(22)\textsuperscript{\emph{b}}\\
               	& 				& 473.23(24)	& 1.12(12)\\
                & 				& 4575.49(24)	& 3.97(28)\\
    4605.7(6) 	& 8(8)			& 158.49(20)  	& 23.3(27) \\   
    4705.6(6) 	& 2.3(13)		& 100.35(29)  	& 2.5(12)\\
    			& 				& 258.16(21)  	& 2.25(24)\\
    4709.6(8) 	& 5.1(22)		& 103.89(24)  	& 5.7(22)\\
	4744.9(6) 	& 0.53(14)		& 297.73(22)  	& 0.53(14)	\\
    4876.0(7) 	& 6.8(10)		& 270.25(20)  	& 9.2(10)\textsuperscript{\emph{a}}\\
	4989.9(6) 	& 0.38(6)		& 432.07(23)  	& 0.38(6)\\
	5221.7(7) 	& \num{\le 0.08}& 516.28(21)    & 1.9(3)\\
	5231.8(5)   & \num{\le 0.07}      & 674.06(28)    & 0.78(9)\\
	5529.2(8) 	& 1.61(19)		& 653.44(28)  	& 0.96(11)\\
    			& 				& 923.4(4)		& 0.63(15)\\
    5543.0(10)  & 0.12(10)      & 1096.1(5)     & 0.44(8) \\
    5579.4(8) 	& 2.73(25)		& 357.56(26)	& 1.25(18)\\
    			& 				& 973.7(3)		& 1.50(16)\\
    5674.1(9) 	& 0.63(25)		& 964.53(25)	& 0.64(25)\textsuperscript{\emph{a}}\\
    6108.6(7)	& 0.30(8)		& 1661.7(6)		& 0.30(8)\\
	6179.9(8) 	& 1.34(17)		& 1621.66(26) 	& 1.34(17)\\
	6266.4(7) 	& 0.50(13)		& 1819.2(4)		& 0.50(13)\\
	6491.8(7)   & 0.25(6)       & 1260.0(5)     & 0.25(6)\\
	6654.0(9) 	& 40.0(15)		& 1111.2(5)     & 0.32(7) \\
	            &               & 1422.0(4)     & 0.61(10) \\
	            &               & 1432.67(29) 	& 1.1(5)\\
    			& 				& 1778.16(29)  	& 1.51(24)\\
                & 				& 1949.0(5)		& 0.58(22)\\
                & 				& 2078.34(23)  	& 5.6(4)\\
               	& 				& 2095.99(20) 	& 19.3(7)\\
                & 				& 2208.0(3)		& 1.7(4)\textsuperscript{\emph{a}}\\
                & 				& 2380.44(20)\textsuperscript{\emph{b}}	& 9.1(4)\textsuperscript{\emph{b}}\\
    6986.9(6) 	& 3.00(23)		& 2381.6(10)\textsuperscript{\emph{b}}    & 0.12(6)\textsuperscript{\emph{b}}\\
    			& 				& 2429.14(24)\textsuperscript{\emph{b}}   & 2.87(22)\textsuperscript{\emph{b}} \\\hline\hline
    \end{tabular}
    \begin{flushleft}
    \textsuperscript{\emph{a}} Corrected for $^{131}$Sn $\beta$-decay contamination.\\
     \textsuperscript{\emph{b}} Measured in coincidence due to doublet\\
        \textsuperscript{\emph{c}} Does not include the additional 8\% uncertainty from $P_{1n}(21/2^+)$ to the total number of $\beta$ decays.\\
    \end{flushleft}
    \label{tab:high_spin_intensities}
\end{center}
\end{table}

The high-spin isomer of $^{131}$In is believed to have the dominant single-particle configuration $\pi g^{-1}_{9/2}\nu f_{7/2}h^{-1}_{11/2}$ and decay by GT $\pi g_{9/2}^{-1}\rightarrow \nu g_{7/2}^{-1}$, and FF ($\pi g_{9/2}^{-1}\rightarrow \nu h_{11/2}^{-1}$ and $\nu f_{7/2}\rightarrow \pi g_{7/2}$) $\beta$ transitions~\cite{PhysRevC.70.034312,bhattacharyya01,wang13}. 
In previous work~\cite{PhysRevC.70.034312}, the $4606+X$-keV isomeric level in $^{131}$Sn was assigned significant direct $\beta$-decay feeding. In Ref.~\cite{bhattacharyya01} this state was suggested to have the configuration of $\nu f_{7/2}h^{-2}_{11/2}$ based on the $B(E2)$ of the de-exciting 158-keV transition compared to the analogous $\nu h_{11/2}^{-2}$ states in $^{130}$Sn. The $\beta$-decay feeding into this level should be somewhat diminished due to the Pauli blocking between the initial $\nu h_{11/2}^{-1}$ hole and the created $\nu h_{11/2}^{-1}$ hole. 
However, only one $\gamma$ ray had previously been observed to feed this level from above~\cite{PhysRevC.70.034312} and its intensity does not appear to have been subtracted when calculating the $\beta$-decay log$ft$ value in Ref.~\cite{PhysRevC.70.034312}. In the current work, using the method described in Fig.~\ref{fig:isomer_timing}, $\gamma$ rays feeding the $23/2^-$ isomer from above were clearly identified, and 5 additional $\gamma$-$\gamma$ coincidences at 100, 104, 923, 974 and 2382~keV were established with the 158-keV $\gamma$ ray. The addition of these $\gamma$ rays to the $^{131}$Sn decay scheme lowers the deduced direct $\beta$-feeding to the $4606+X$-keV level to a value of 8(8)\% which is consistent with zero, although a large uncertainty remains due to the unknown internal conversion coefficients of the 100-, 104- and 158-keV transitions. Measuring the electron conversion coefficients of these low energy transitions would greatly reduce this uncertainty. 

The strong $\beta$-decay transition to the $6654+X$-keV level was previously assigned as an allowed GT decay of $\pi g^{-1}_{9/2}\rightarrow \nu g^{-1}_{7/2}$~\cite{PhysRevC.70.034312}. The assignment was based purely on similarities between this decay and the strength of the $\beta$ decay of the $9/2^+$ ground state of $^{131}$In via the analogous $\pi g_{9/2}^{-1}\rightarrow \nu g_{7/2}^{-1}$ decay. In fact, the log$ft$ values for these decays are very similar: 4.4 and 4.6 for the 2434-keV and $6654+X$-keV levels, respectively. The $\beta$-decay transition to the $6987+X$-keV level is also relatively strong (log$ft\sim5.6$), and shows a similar decay pattern as the $6654+X$-keV state. This state also has an energy consistent with an assignment of a $21/2^+$ state with the configuration of $\nu f_{7/2} g_{7/2}^{-1}h_{11/2}^{-1}$ at 7.018-MeV excitation energy predicted in Ref.~\cite{wang13}.

In the current experiment, a set of positive-parity states at $4246+X$-, $4424+X$-, $4575+X$- and $4990+X$-keV that were previously proposed to have the configuration $\nu f_{7/2}d^{-1}_{3/2}h^{-1}_{11/2}$~\cite{bhattacharyya01} were observed to be populated following $\beta$ decay of $^{131}$In for the first time. These states were previously observed via fission of $^{248}$Cm with \mbox{Gammasphere}~\cite{bhattacharyya01}. However, the apparent direct $\beta$-decay feeding to some of these states would correspond to the $\pi g_{9/2}^{-1}\rightarrow \nu d_{3/2}^{-1}$ $\beta$ decay, which would be second-forbidden unique, and therefore, highly suppressed. This suggests that there is either unobserved $\gamma$-ray feeding from above, or these states have significant mixing with other configurations that result in allowed $\beta$ transitions.

The $21/2^-$ member of the $\nu f_{7/2}h_{11/2}^{-2}$ multiplet is predicted to be close in energy to the $23/2^-$ state~\cite{bhattacharyya01,wang13}. The 100 and 258~keV transitions from the $4706+X$-keV level compete with each other suggesting that these transitions are of similar multipolarity. This supports the assignment of the $4706+X$-keV state as the $(21/2^-)$ member of the $\nu f_{7/2}h_{11/2}^{-2}$ multiplet decaying to both the $(19/2^-)$ and $(23/2^-)$ states of this multiplet via $M1$ transitions. Furthermore, this would suggest that the 1949-keV transition is a weak $E1$ transition from the $6654+X$-keV $(21/2^+)$ state. 

\begin{figure}[!t]
   \includegraphics[width=\linewidth]{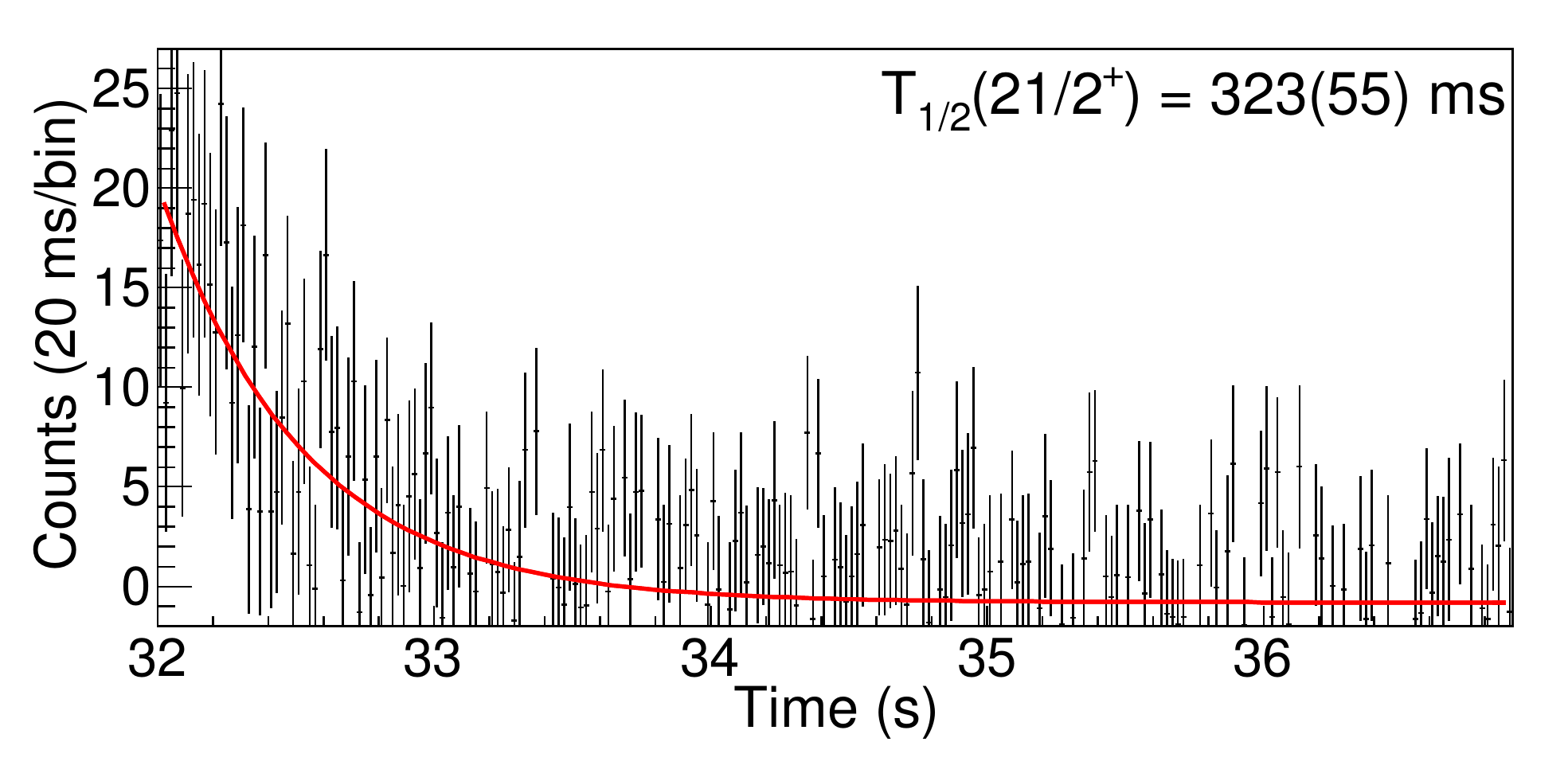}%
   \caption{(Color online) Measurement of the half-life of the $21/2^+$ isomer in $^{131}$In. The activity curve from the ``beam-off'' portion of the cycle was generated by placing a gate on the 171-, 173- and 285-keV $\gamma$ rays, which are emitted following the $\beta$ decay of the $21/2^+$ state. The half-life from this fit was measured to be 323(55)~ms.
   \label{fig:21-2-halflife}}
\end{figure}

Finally, the half-life of the $21/2^+$ $\beta$-decaying isomer of $^{131}$In was measured by fitting the time-dependent activity of the 171- , 173- and 285-keV $\gamma$ rays in the ``beam-off'' portion of the cycle. As shown in Fig.~\ref{fig:21-2-halflife}, the half-life of this state was measured to be 323(55)~ms, in good agreement with the value of 320(60)~ms reported in Ref.~\cite{PhysRevC.70.034312}.

\subsection{Measurement of the total number of $\beta$ decays for each state of $^{131}$In}\label{sec:number_b_decays}
In order to determine the absolute intensity of each $\gamma$-ray per $\beta$ decay, the total number of observed decays for each $\beta$-decaying state in $^{131}$In is required. While a $\beta$-decay transition from the high-spin $21/2^+$ isomer in $^{131}$In is always followed by a relatively high-energy $\gamma$ ray, the lower-lying isomers of $^{131}$In can $\beta$ decay directly to states in $^{131}$Sn that do not emit $\gamma$ rays.

\begin{figure}[t!]
   \includegraphics[width=\linewidth]{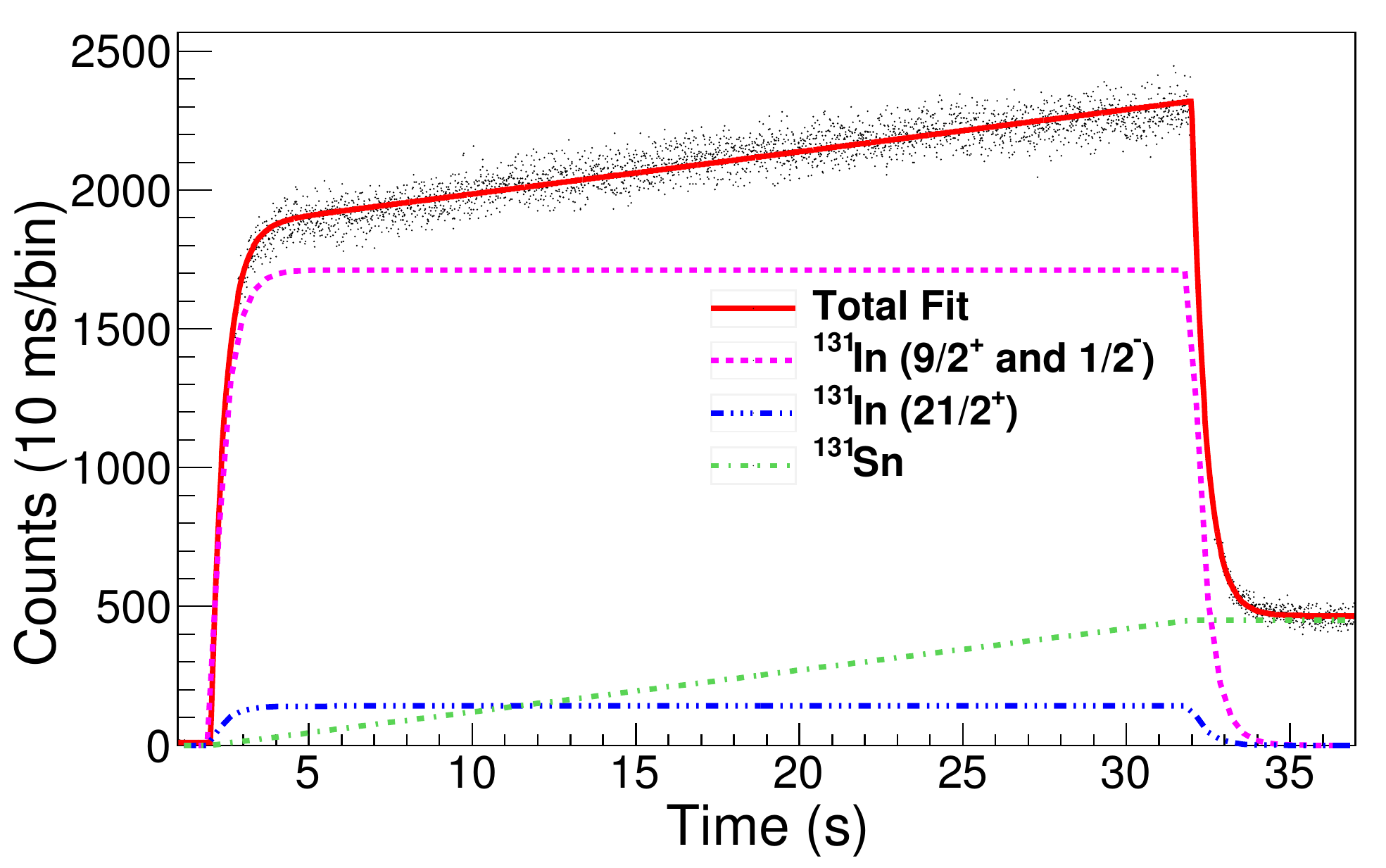}%
   \caption{(Color online) Fit of the SCEPTAR $\beta$ detector activity summed over all 182 cycles. The intensity of the high-spin $^{131}$In decay was fixed to reproduce the total measured $\beta$ decay intensity into the high-spin states of $^{131}$Sn as described in Section~\ref{sec:high-spin}.
   \label{fig:beta_cycle_fit}}
\end{figure}

The total number of $\beta$ decays of $^{131}$In was determined using a fit of the time structure of the $\beta$ particles measured in SCEPTAR (Fig.~\ref{fig:beta_cycle_fit}). The contribution from the high-spin $^{131}$In decay was fixed to the value of $4.10(33)\times10^5$ events measured from the $\gamma$-ray intensities in the high-spin portion of the level scheme as described in Section~\ref{sec:high-spin}. A constant background rate was used, as well as a component for the decay of the daughter $^{131}$Sn, including the $^{131}$Sn in the beam. Additionally, the $\beta n$ decay of $^{131}$In was included in order to reproduce both the measured $\beta n$ events, as well as the subsequent decays of the $^{130}$Sn nuclei.
From the combined fit, shown in Fig.~\ref{fig:beta_cycle_fit}, the number of $\beta$ decays from the low-spin states of $^{131}$In was measured to be $4.897(40)\times10^6$, including the contribution from the mixed $9/2^+$ and $1/2^-$ $\beta n$ branch, as well as the uncertainty in $P_{1n}(21/2^+)$. 

\subsection{$\beta$-Decay scheme of the $9/2^+$ state}
The $\pi g^{-1}_{9/2}$ $9/2^+$ single-particle proton-hole state forms the ground state of $^{131}$In. A strong allowed $\beta$-transition is expected to the $\nu g^{-1}_{7/2}$ $7/2^+$ state of $^{131}$Sn, followed by a strong $E2$ $\gamma$-ray transition to the $\nu d_{3/2}^{-1}$ $3/2^+$ ground state of $^{131}$Sn. The only other potentially significant $\beta$-fed low-lying level in $^{131}$Sn is the FF $\beta$-decay to the $\nu h^{-1}_{11/2}$ $11/2^-$ state.

\begin{figure}[t!]
   \includegraphics[width=\linewidth]{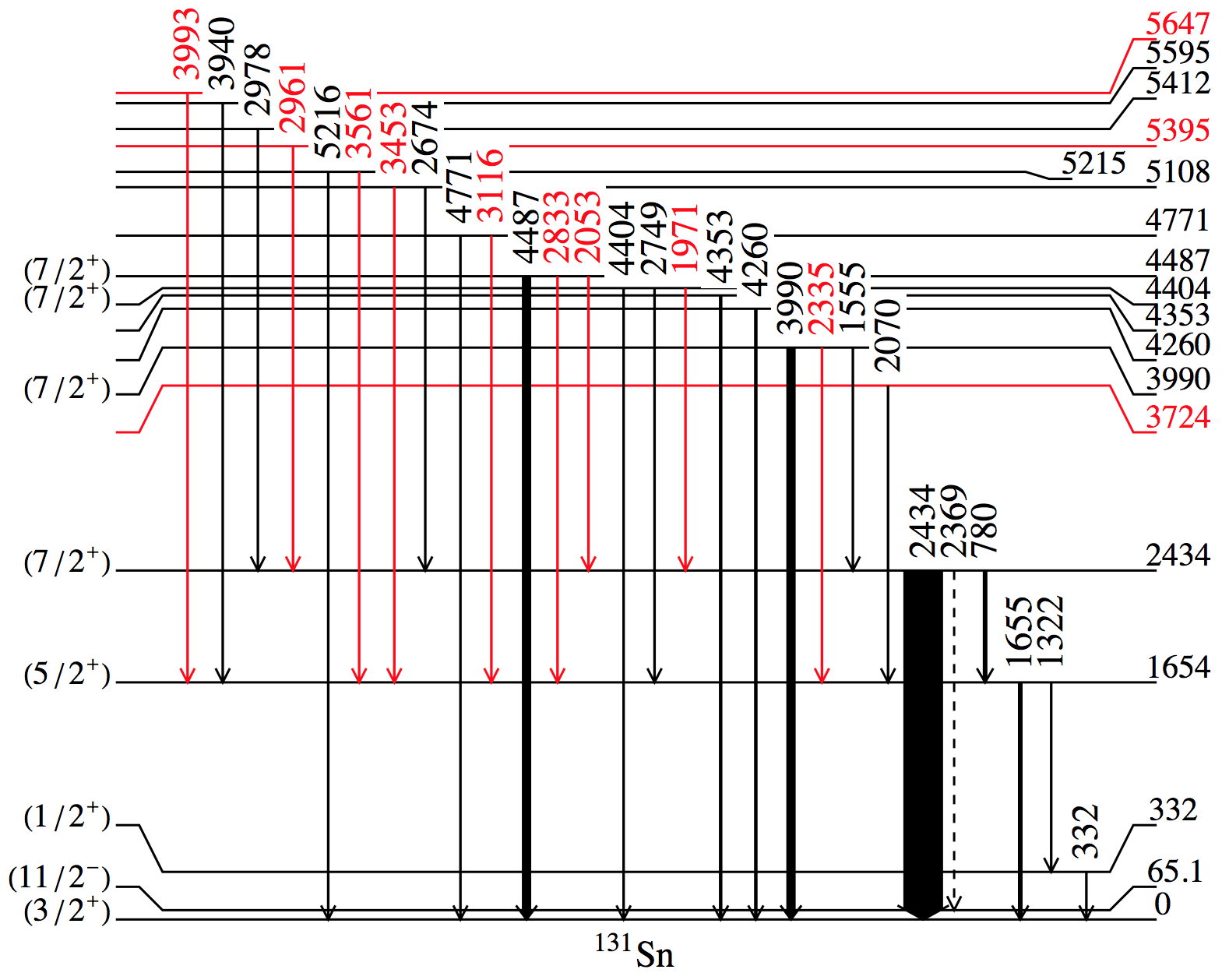}%
   \caption{(Color Online) The decay scheme of $^{131}$Sn observed following the $\beta$ decay of the $9/2^+$ isomeric state in $^{131}$In. The widths of the arrows represents the relative intensities of each of the transitions. Previously unobserved transitions and excited states are highlighted in red. The location of the 2369~keV $\gamma$-ray transition could not be confirmed. 
   \label{fig:sn131-med-spin}}
\end{figure}

The low-lying 332~keV excited state in $^{131}$Sn is the $\nu s_{1/2}^{-1}$ $1/2^+$ state. In this work, it is assumed that any high-lying states in $^{131}$Sn that have a direct $\gamma$-ray transition to this state --- with no observed $\gamma$-ray feeding from above --- originate from the $\beta$-decay of the $1/2^-$ state of $^{131}$In. For a FF $\beta$ transition from the $9/2^+$ state, the daughter states have a minimum spin of $5/2^-$, meaning that the lowest multipolarity for a direct $\gamma$-ray transition to the $1/2^+$ state at 332-keV excitation energy would be $M2/E3$, which, in general, would not compete with a transition of $E1, M1$, or $E2$ character to one of the other low-lying states of $^{131}$Sn. Moreover, the first $5/2^-$ state in $^{131}$Sn is predicted to be at an energy of 4.7~MeV~\cite{wang13}, while Ref.~\cite{kozub12} has tentatively assigned a $5/2^-$ state from the 1p-2h $f_{5/2}$ excitation at 4655(50)~keV. No excited state measured in the current work is consistent with a $f_{5/2}(2h)$ state at that energy, which would, however, be difficult to populate in $\beta$ decay. The assumption that a state that $\gamma$ decays directly to the $1/2^+$ state at 332-keV excitation energy is not directly fed by the $\beta$-decay of the $9/2^+$ in $^{131}$In does, however, ignore potential contributions to the population of these states from the ``Pandemonium Effect''~\cite{Hardy77}.
The assignment of the decay strengths and spin-parities of these excited states should thus be considered tentative. The rest of the high-lying levels that bypass the 332-keV $1/2^+$ state are assigned to the decay scheme of the $9/2^+$ isomer. With these criteria, 3 new excited states were observed following this decay, including 9 new $\gamma$-ray transitions. Figure~\ref{fig:sn131-med-spin} shows the determined decay scheme while Tab.~\ref{tab:med_spin_intensities} shows the measured intensities of each $\gamma$ ray following the $\beta$-decay of the $9/2^+$ state of $^{131}$In.

The total number of $\beta$ decays observed to feed $\gamma$-decaying states in $^{131}$Sn following the $\beta$ decay of the $9/2^+$ state in $^{131}$In was $2.36(4)\times10^6$. The direct $\beta$ decay from the $9/2^+$ ground state of $^{131}$In to the $3/2^+$ ground-state of $^{131}$Sn is negligible due to the $\beta$ decay being second forbidden. However, the FF transition to the $11/2^-$ state at 65~keV in $^{131}$Sn may have a non-negligible branch. This work follows the same procedure as Ref.~\cite{PhysRevC.70.034312} and uses the measured $\beta$-decay strengths in $^{127,129}$In decays to estimate the direct $\beta$-feeding to this state from the $9/2^+$ ground state of $^{131}$In to be approximately 5\%. The possibility that the $^{127,129}$In FF $\beta$-decay strength may be systematically low due to potential Pauli blocking of the hole states~\cite{PhysRevC.70.034312,zhi13} is considered, and an uncertainty is assigned to allow for up to a 20\% $\beta$ branch to this state from the $9/2^+$ ground state of $^{131}$In. Using a $\beta$ feeding of $5^{+15}_{-5}\%$ and including the $\beta n$ feeding of 2(2)\% results in the total number of $\beta$ decays from the $9/2^+$ ground state of $^{131}$In observed in this experiment to be $2.53^{+49}_{-13}\times10^6$, implies that $2.36^{+13}_{-49}\times10^6$ decays were observed from the $1/2^-$ state of $^{131}$In.

The weak $E2$ transition between the $5/2^+$ and $1/2^+$ states in $^{131}$Sn was observed via the coincidence between the 780-keV and 1322-keV $\gamma$ rays. This allowed the measurement of the branching ratio for the 1322-keV $E2$ transition from the 1654-keV state and the removal of the upper limit~\cite{PhysRevC.70.034312} for the strength of this transition.

\begin{table}[!t]
\begin{center}
    \caption{The observed excited states and $\gamma$-ray transitions in $^{131}$Sn following the $\beta$-decay of the $9/2^+$ ground state in $^{131}$In. For the $\gamma$ decaying states, $I_{out}-I_{in}$ represents the excess observed $\gamma$ ray and conversion electron intensity out of a state compared to the intensity into the state. This represents an upper limit for the $\beta$ feeding.} 
    \sisetup{table-align-text-post = false}
    \begin{tabular}{  l  S[table-format=2.5]  S[table-format=5.5] S[table-format=3.5] S[table-format=4.6]}
    \hline
    \multicolumn{1}{c}{\multirow{3}{*}{\shortstack[c]{Level\\Energy\\(keV)}}}	& & \multicolumn{1}{c}{\multirow{3}{*}{\shortstack[c]   {$\gamma$-ray\\Energy\\(keV)}}} & \multicolumn{1}{c}{\multirow{3}{*}{\shortstack[c]{Absolute\\Intensity\\(\%)$^c$}}} & \multicolumn{1}{c}{\multirow{2}{*}{\shortstack[c]{Relative\\Intensity\\(\%)}}}  \\ 
    
    & \multicolumn{1}{c}{\multirow{2}{*}{\shortstack[c]{$I_{out}$ - $I_{in}$\\(\%)$^c$}}} &  & \\	
    
   & & &					\\
    \hline\hline
    $X$				&	5\textsuperscript{\emph{a}}		&				&			&\\ 
    331.72(20)	  	&			& 331.72(20)\textsuperscript{\emph{b}} 	& 0.040(15)       & 0.049(19)\\
    1654.48(20) 	&			& 1322.2(5)		& 0.040(15) 	& 0.049(19)\\
    				& 			& 1654.53(20)  	& 2.18(10) 	& 2.64(12)\\
   	2434.1(4) 	 	& 84.2(22)  & 779.52(20)  	& 1.69(6) 		& 2.04(8)\\
    				&  			& 2369.0(3)		& 0.15(3) 		& 0.18(4)\\
                    &			& 2434.1(5)\textsuperscript{\emph{b}} 	& 82.6(15) 		& 100.0(11)\\ 
    3724.0(5)		& 0.022(9) & 2069.5(5)		& 0.022(9) 		& 0.027(12)\\
    3989.78(20) 	& 3.09(15)	& 1555.2(5) 	& 0.042(15) 	& 0.051(18)\\  
    				&		 	& 2335.0(4)		& 0.12(4)		& 0.15(4)\\
                    &           & 3989.86(20)	& 2.92(14) 		& 3.54(16)\\
    4259.8(5)		& 0.13(5)	& 4259.8(5)		& 0.13(5) 		& 0.15(6)\\
    4353.2(7)		& 0.060(29)	& 4353.2(7)		& 0.060(29) 	& 0.07(3)\\
    4404.1(4)		& 0.30(6) 	& 1971.3(8)  	& 0.072(25)		& 0.09(3)\\ 
    				&			& 2749.1(3)  	& 0.055(14) 	& 0.067(17)\\
                    &           & 4404.2(4)  	& 0.17(6) 		& 0.21(7)\\
    4487.25(21) 	& 4.06(16)	& 2053.3(4)  	& 0.037(17)		& 0.044(20)\\
    				&			& 2833.3(5)  	& 0.064(13) 	& 0.077(16)\\
                    &           & 4487.17(21)	& 3.97(16)		& 4.80(18)\\
    4770.8(4)   	& 0.65(4) 	& 3115.7(6) 	& 0.062(15)		& 0.076(18)\\
    				&			& 4770.9(4)		& 0.59(4)		& 0.71(5)\\
    5107.7(12)  	& 0.024(13) & 2673.8(7)  	& 0.015(10) 	& 0.018(12)\\
    				&			& 3453.0(7)		& 0.009(8) 		& 0.011(9)\\
	5215.8(5)		& 0.26(3)	& 3560.5(12)	& 0.020(11) 	& 0.024(14)\\
    				&			& 5216.0(5)		& 0.24(3) 		& 0.28(4)\\
    5394.6(11)  	& 0.040(22) & 2960.5(10)  	& 0.040(22) 	& 0.048(27)\\         
    5412.4(12)     & 0.041(21) & 2978.2(11)    & 0.041(21)		& 0.050(25)\\
    5594.7(9) 		& 0.07(4)	& 3940.2(8)   	& 0.07(4)		& 0.08(4)\\
    5647.3(9)	 	& 0.11(4)	& 3992.7(8)\textsuperscript{\emph{b}}  & 0.11(4)\textsuperscript{\emph{b}}   & 0.13(4)\\
 \hline\hline
    \end{tabular}
    \begin{flushleft}
    \textsuperscript{\emph{a}} Corrected for $^{131}$Sn $\beta$-decay contamination\\
    \textsuperscript{\emph{b}} Measured in coincidence due to doublet\\
   \textsuperscript{\emph{c}} Does not include the additional -15\% and +5\% uncertainty in the FF $\beta$ feeding to the $11/2^-$ state from $9/2^+$ decay.\\
    \end{flushleft}
    \label{tab:med_spin_intensities}
\end{center}
\end{table}

The levels at 3990-, 4404- and 4487-keV excitation energy were all observed to decay to each of the low-lying $3/2^+$, $5/2^+$ and $7/2^+$ states, with no observed intensity to the low-lying $1/2^+$ or $11/2^-$ states. The 3990- and 4487-keV levels also have reasonably large $\beta$ feeding with log$ft \approx 5$. This decay pattern is consistent with these levels having a spin-parity of $(7/2^+)$. The predicted energies for two of the excited $7/2^+$ states in $^{131}$Sn from the calculations of Ref.~\cite{wang13} are 3923~keV and 4553~keV, in very good agreement with this assignment. It should be noted that the observed levels at 3724~keV, 4260~keV, 4353~keV and each level above $> 4.5$~MeV may have been populated by the $\beta$ decay of the $1/2^-$ isomer in $^{131}$In. However, the total contribution of these levels to the decay of both the $9/2^+$ $\beta$-decay intensity is less than 1.4\%, which is insignificant compared to the uncertainty in the FF decay to $11/2^-$ state.

\begin{figure}[!t]
   \includegraphics[width=\linewidth]{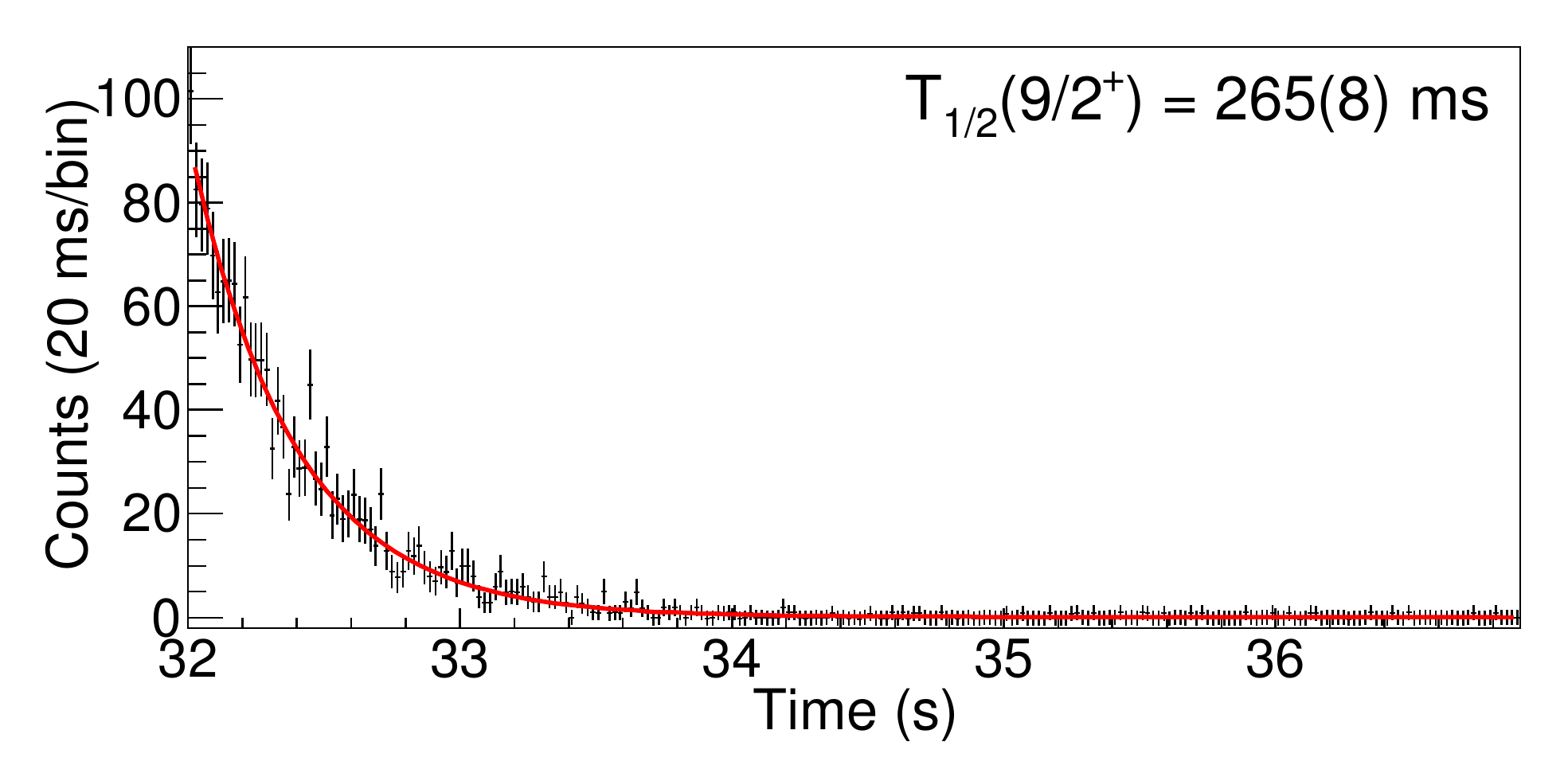}%
   \caption{Fit of the half-life of the $9/2^+$ ground state of $^{131}$In. The activity curve shown was generated by placing a gate on the 2434~keV $\gamma$ ray which is emitted primarily following the $\beta$ decay of the $9/2^+$ state. A narrow gate was chosen in order to exclude the contribution from the 2429~keV $\gamma$-ray transition from the $21/2^+$ isomeric decay. The half-life from this fit was measured to be 265(8)~ms.
   \label{fig:halflife-2434}}
\end{figure}

The state at 2434-keV excitation energy in $^{131}$Sn with a spin-parity of $7/2^+$ was assumed to be populated mainly from the decay of the $9/2^+$ state of $^{131}$In. In principle, there could be weak $\gamma$-ray feeding into this state from above following the FF $\beta$-decay of the $1/2^-$ state of $^{131}$In to high-lying excited $3/2^+$ and $5/2^+$ states of $^{131}$Sn. However, the majority of the observed population is clearly from direct $\beta$ feeding via the allowed $\pi g_{9/2}^{-1}\rightarrow \nu g_{7/2}^{-1}$ decay of the $9/2^+$ ground state of $^{131}$In. Figure~\ref{fig:halflife-2434} shows the fit of the time dependence of the 2434-keV $\gamma$ ray during the decay portion of the cycle with a gate between 2432~keV and 2437~keV. The half-life was measured to be 265(8)~ms, in good agreement with the most precise previous measurement of 261(3)~ms~\cite{lorusso15}. A chop analysis was performed on this measurement and did not show any systematic rate-dependent effects.

It should be noted that in both the $21/2^+$ and $9/2^+$ decays, there were $\gamma$-rays emitted from states above 
the neutron separation energy of 5204(4)~keV. This has also been observed for neighboring isotopes~\cite{jungclaus16} and seems to be a common effect for many neutron-rich nuclei. These $\gamma$ rays are in 
competition with neutron emission, and theoretical models have neglected them for a long time. Recent developments of QRPA-models in conjunction with Hauser-Feshbach descriptions~\cite{mumpower162,moller19} describe this competition much better and can be benchmarked with data from highly-efficient setups like GRIFFIN.

\subsection{$\beta$-Decay scheme of the $^{131}$In $1/2^-$ isomer}
\begin{figure}[t!]
   \includegraphics[width=0.7\linewidth]{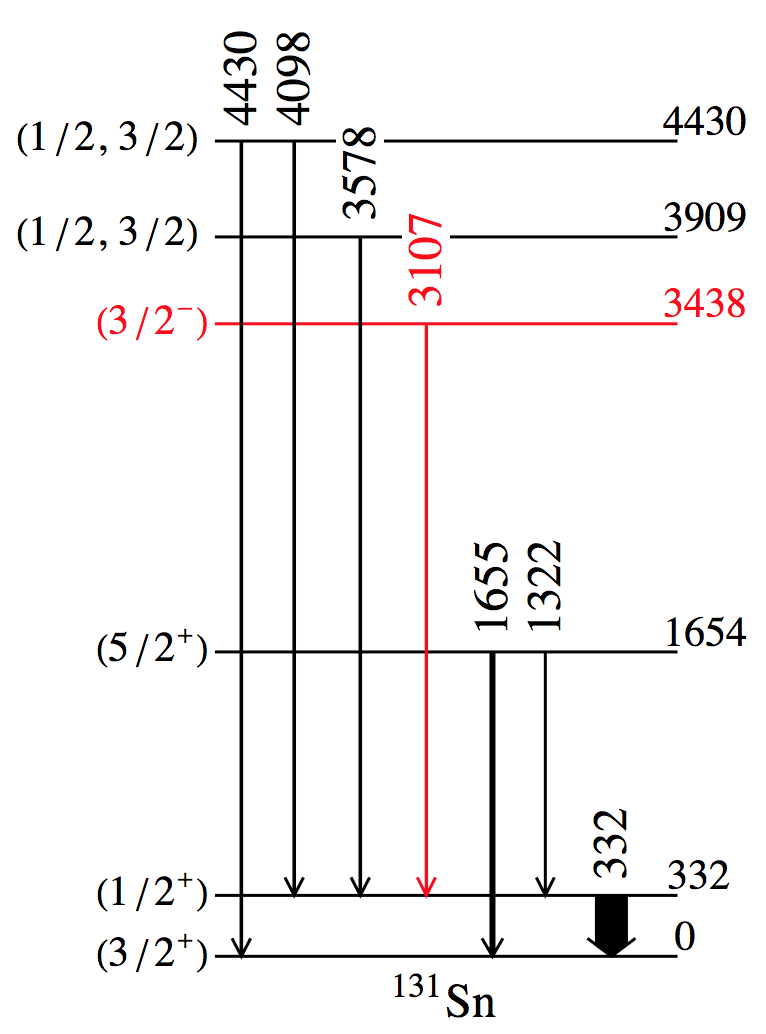}%
   \caption{(Color Online) Decay scheme of $^{131}$Sn observed following the $\beta$ decay of the $1/2^-$ state in $^{131}$In. The widths of the arrows represents the relative intensities of each of the transitions. Previously unobserved transitions and excited states are highlighted in red.
   \label{fig:sn131-low-spin}}
\end{figure}

The $\beta$-decaying $1/2^-$ isomeric state in $^{131}$In is believed to be characteristic of the $\pi p^{-1}_{1/2}$ single-particle proton-hole state. There are no $1/2^-$ or $3/2^-$ states at low-excitation energy in $^{131}$Sn that could be populated by allowed $\beta$-decays. Therefore, one would expect a significant fraction of the $\beta$-decay strength from this state to be in FF transitions to the positive-parity $\nu d^{-1}_{3/2}$ ground state and the $\nu s^{-1}_{1/2}$ and $\nu d^{-1}_{5/2}$ low-lying excited states of $^{131}$Sn to be in competition with allowed $\beta$-decays to $1/2^-$ and $3/2^-$ states at higher excitation energies. As shown in Fig.~\ref{fig:sn131-low-spin} and Tab.~\ref{tab:low_spin_intensities}, this is indeed what was observed in the current work.

\begin{figure}[!t]
   \includegraphics[width=\linewidth]{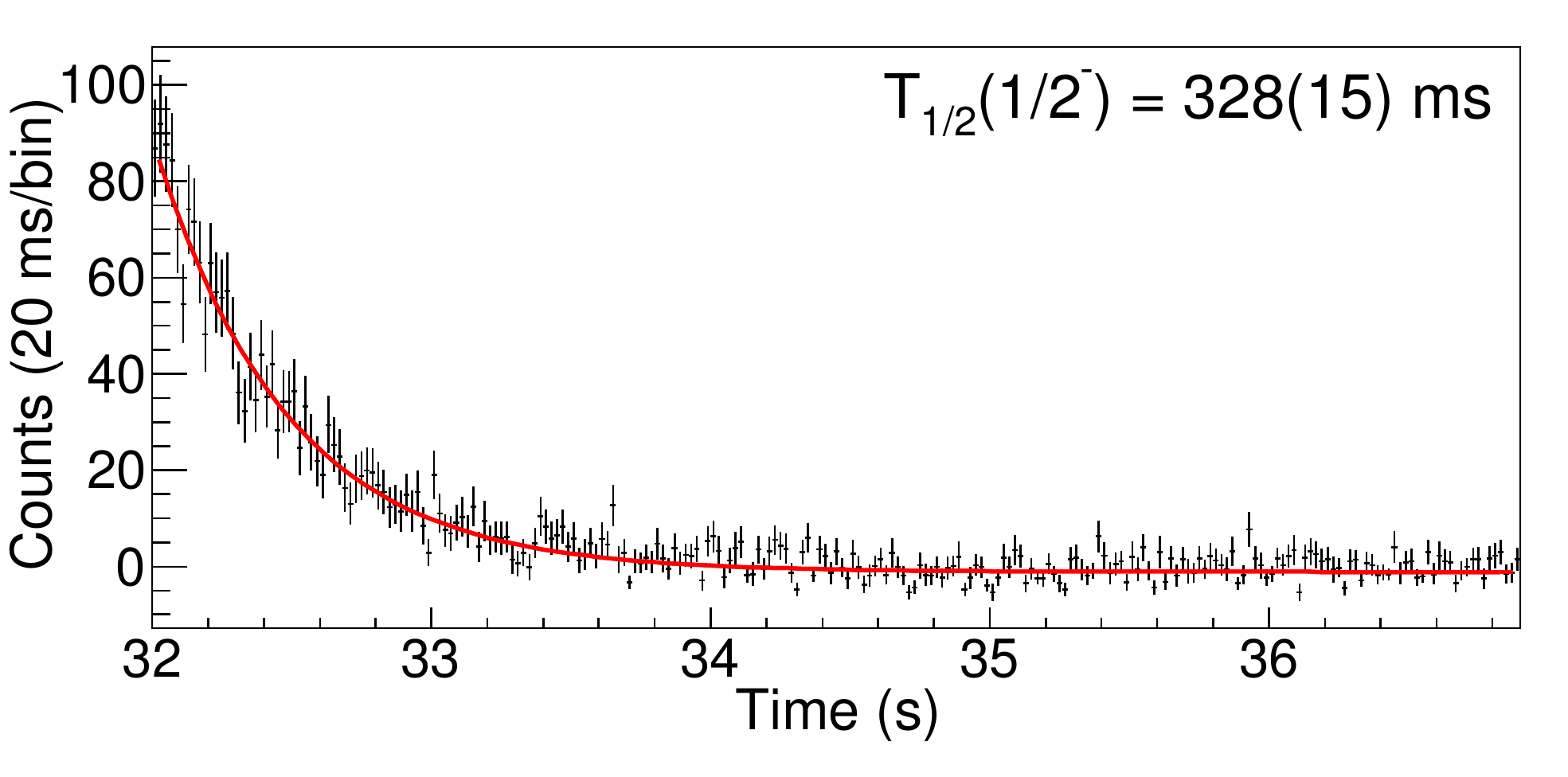}%
   \caption{(Color Online) Fit of the half-life of the $1/2^-$ isomer in $^{131}$In. The activity curve shown was generated by placing a gate on the 332~keV $\gamma$ ray which is emitted primarily following the $\beta$ decay of the $1/2^-$ state. The feeding from the $\beta$ decays of the other $^{131}$In states was negligible. The 332~keV $\gamma$ decay in $^{131}$Sb following the $\beta$ decay of the $^{131}$Sn daughter was also accounted for. The half-life from this fit was measured to be 328(15)~ms. 
   \label{fig:1-2-halflife}}
\end{figure}

The half-life of the $1/2^-$ isomer was measured by fitting the time-dependence of the 332-keV $\gamma$ ray. The relative intensity of the 330-keV $\gamma$ ray from the decay of the $21/2^+$ state was determined to be $< 1\%$ of the 332-keV $\gamma$ ray. Both the contribution of the $9/2^+$ decay that proceeds through the 332-keV $\gamma$-ray transition via the weak 1322-keV $E2$ branch from the $5/2^+$ level at an excitation energy of 1654~keV, as well as the 330~keV transition, were accounted for in the fit and had only a small effect on the measured half life. The half-life fit also took into account the small contribution of the 332~keV $\gamma$-ray emitted following the $\beta$ decay of $^{131}$Sn. As shown in Fig.~\ref{fig:1-2-halflife}, the half-life of the $1/2^-$ state was measured to be 328(15)~ms, in good agreement with, but a factor of 3 more precise than, the value of 350(50) reported in Ref.~\cite{PhysRevC.70.034312}.

In this work, a new level at 3438~keV was placed in the $^{131}$Sn decay scheme based on a coincidence of 332-3107-keV transitions. This state is a potential candidate for the 3404(50)-keV $\nu p_{3/2}(2h)$ $(3/2^-)$ state reported in Ref.~\cite{kozub12}. This state has also been observed in the $^{130}$Sn($^9$Be,$^8$Be)$^{131}$Sn reaction~\cite{JonesPriv}. Although a $1/2^-\rightarrow3/2^-$ decay would be an allowed $\beta$-transition from spin-parity considerations, the matrix element requires 2p-2h mixing across the $N=82$ shell gap, and is therefore observed to be strongly suppressed with a log$ft$ of 7.4, consistent with a robust $N=82$ shell closure in $^{131}$In.

\begin{table}[!t]
\begin{center}
    \caption{The observed excited states and $\gamma$-ray transitions in $^{131}$Sn following the $\beta$-decay of the $1/2^-$ isomer in $^{131}$In. For the $\gamma$ decaying states, $I_{out}-I_{in}$ represents the excess observed $\gamma$ ray and conversion electron intensity out of a state compared to the intensity into the state. This represents an upper limit for the $\beta$ feeding.} 
    \sisetup{table-align-text-post = false}
    \begin{tabular}{  l  S[table-format=2.5]  S[table-format=5.5]  S[table-format=3.5] S[table-format=4.5] }
    \hline
    \multicolumn{1}{c}{\multirow{3}{*}{\shortstack[c]{Level\\Energy\\(keV)}}}	& & \multicolumn{1}{c}{\multirow{3}{*}{\shortstack[c]   {$\gamma$-ray\\Energy\\(keV)}}} & \multicolumn{1}{c}{\multirow{3}{*}{\shortstack[c]{Absolute\\Intensity\\(\%)$^c$}}} & \multicolumn{1}{c}{\multirow{2}{*}{\shortstack[c]{Relative\\Intensity\\(\%)}}}  \\ 
    
    & \multicolumn{1}{c}{\multirow{2}{*}{\shortstack[c]{$I_{out}$ - $I_{in}$\\(\%)$^c$}}} &  & \\	
    
   & & &					\\
    \hline\hline
    0				& 68.1(22)	&			 	& 		 		&\\
    331.72(20)  	& 27.8(7)	& 331.72(20)\textsuperscript{\emph{b}} 	&                     28.6(6)\textsuperscript{\emph{a}} 	                        & 100.0(14)\\
    1654.48(20)		& 0.68(3)	& 1322.2(5)  	& 0.012(5) 		& 0.043(16)\\
    				&			& 1654.53(20)	& 0.67(3) 		& 2.33(10)\\
    3438.3(5)		& 0.071(17) & 3106.6(3) 	& 0.070(17)\textsuperscript{\emph{a}}	                        & 0.25(6)\\
    3909.2(5) 		& 0.210(29) & 3577.5(4)  	& 0.210(29)		& 0.73(10)\\
    4430.0(3)   	& 1.17(8)	& 4098.4(3)\textsuperscript{\emph{b}} 	& 0.53(5)\textsuperscript{\emph{b}}		& 1.86(16)\\
    				&			& 4429.95(26) 	& 0.63(6)		& 2.11(121)\\
 \hline\hline
    \end{tabular}
    \begin{flushleft}
    \textsuperscript{\emph{a}} Corrected for $^{131}$Sn $\beta$-decay contamination\\
        \textsuperscript{\emph{b}} Measured in coincidence due to doublet\\
    \textsuperscript{\emph{c}} Does not include the additional -15\% and +5\% uncertainty in the FF $\beta$ feeding to the $11/2^-$ state from $9/2^+$ decay.\\
\end{flushleft}
    \label{tab:low_spin_intensities}
\end{center}
\end{table}
The total number of $\beta$ decays observed to feed $\gamma$-decaying states in $^{131}$Sn following the $\beta$ decay of the $1/2^-$ state in $^{131}$In was $7.07(9)\times10^5$, not including the 2(2)\% $\beta n$ decay branch. An important consequence of the more detailed spectroscopy of $^{131}$Sn performed in the current work is a significant revision of the $\beta$-decay branching ratios for the decay of the $1/2^-$ isomer in $^{131}$In compared to previous works~\cite{PhysRevC.70.034312,fogelberg84}. In Ref.~\cite{PhysRevC.70.034312}, the ground-state $\beta$-feeding from the $1/2^-$ state of $^{131}$In was determined to be 95\%, while the $\beta$ feeding to the 332-keV excited state was reported to be 3.5(9)\%, compared to 68.1(22)\% and 27.8(7)\% measured in the current work, respectively. Removing all potential FF $\beta$-feeding of the $11/2^-$ in $^{131}$Sn from the $\beta$ decay of the $9/2^+$ state in $^{131}$In would only change these $\beta$ branches to 69.6(22)\% and 26.4(7)\%, respectively. The authors of Ref.~\cite{PhysRevC.70.034312} adopt the intensity of the 332-keV $\gamma$-ray transition from the measurement performed in Ref.~\cite{fogelberg84}. In Ref.~\cite{fogelberg84}, however, the neglect of the Pandemonium effect~\cite{Hardy77} resulted in many of the $\beta$ decays from $^{131}$In being assigned to the direct ground-state $\beta$-transition from the $1/2^-$ state. However, many of these $\beta$ decays are actually weak transitions to excited states which subsequently $\gamma$-decay to the low-lying states. The majority of this discrepancy can be reconciled simply by considering that when the 3.6\% intensity of the 332-keV transition in Ref.~\cite{fogelberg84} was measured, they also deduced a $\gamma$-ray intensity for the 4273-keV transition of 100\%, and a $\gamma$-ray intensity for the 2434-keV transition of 93\%. In Ref.~\cite{PhysRevC.70.034312}, the observation of additional $\beta$ strength to newly observed states reduced the intensities reported for these transitions to 87\% and 90\%, respectively. If that intensity was accounted for, the total number of $\beta$ decays assigned to the $1/2^-$ state in $^{131}$In would have been reduced, which, in turn, would have decreased the deduced $\beta$-branch to the $3/2^+$ ground state of $^{131}$Sn while increasing the $\beta$ branch to the 332-keV state. Therefore, using the 332-keV intensity per $\beta$ decay reported in Ref.~\cite{fogelberg84} to determine branching ratios for the measurement performed in Ref.~\cite{PhysRevC.70.034312} is incorrect and is likely the primary source of this discrepancy with the current work.

The increased feeding into the 332-keV state deduced in the current work reduces the log$ft$ value for this decay from 6.5 to 5.5, in better agreement with the value of 5.74 predicted in Ref.~\cite{zhi13}, and comparable to the ground state log$ft$ of 5.2. It is important to note that these two transitions in the decay of the $1/2^-$ isomer in $^{131}$In are used in Ref.~\cite{zhi13} to determine the quenching factors for the FF operators in half-life calculations in the region. The reduced quenching of the FF feeding implied by the current work would, however, be expected to result in an even shorter calculated half-life for $^{131}$In. More investigation into the discrepancy between the calculated half-lives in $^{131}$In decay is required.

In addition to the $\gamma$-ray transitions shown in Figs.~\ref{fig:sn131-high-spin},~\ref{fig:sn131-med-spin} and \ref{fig:sn131-low-spin}, there were also a number of high-energy $\gamma$ rays that were observed in the current experiment that could not be placed into the decay scheme based on energy differences or $\gamma$-$\gamma$ coincidences. These $\gamma$ rays are listed in Table~\ref{tab:tentatives} with intensities relative to the strongest $\gamma$ rays from each of the lower-spin decay schemes. These $\gamma$ rays all had grow in and decay time structures consistent with the decay of $^{131}$In. Due to the relatively large photopeak efficiency at 332 and 1654~keV of 20.82(26)\% and 11.11(14)\%, respectively, and the number of counts in singles for the $\gamma$ rays in Tab.~\ref{tab:tentatives}, the lack of observed coincidences implies that these $\gamma$ rays do not feed either of these states. This implies that these are direct $\gamma$-ray transitions to either the $3/2^+$ ground state or $11/2^-$ isomeric state. More detailed spectroscopic information is required for definitive placement of these $\gamma$ rays in the level scheme.
\begin{table}[!t]
\begin{center}
    \caption{Observed $\gamma$-ray transitions that could not be placed in the level scheme of $^{131}$Sn. These transitions had a consistent time structure for the decay of $^{131}$Sn. No coincidences or level energy differences for these $\gamma$ rays were observed. These could be direct transitions to either the $3/2^+$ ground state or $11/2^-$ isomeric state following the $\beta$ decay of $^{131}$In.} 
    \sisetup{table-align-text-post = false}
    \begin{tabular}{  l  S[table-format=2.6]  S[table-format=2.6] }
    \hline
    \multicolumn{1}{c}{$\gamma$-ray}	& \multicolumn{1}{c}{} & \multicolumn{1}{c}{} \\ 
    
    \multicolumn{1}{c}{Energy} & \multicolumn{1}{c}{$I_{Rel}$(332)}	& \multicolumn{1}{c}{$I_{Rel}$(2434)}	\\	
    
    \multicolumn{1}{c}{(keV)} & \multicolumn{1}{c}{(\%)} & \multicolumn{1}{c}{(\%)}										\\
    \hline\hline
    4349.4(3)	& 0.23(9)	& 0.074(28)	\\
    5004.9(7)  	& 0.15(4)	& 0.050(14)	\\
    6340.8(8)	& 1.47(29)	& 0.48(10)	\\
    6553.8(6)  	& 2.4(4)	& 0.78(13) 	\\
    6796.5(10)	& 1.6(3)	& 0.52(11)	\\
    6866.9(7) 	& 2.2(4)	& 0.70(12)	\\ 	
    7086.0(9)   & 1.2(3)	& 0.38(10)	\\  
 \hline\hline
    \end{tabular}
    \label{tab:tentatives}
\end{center}
\end{table}

\section{Conclusion}
The level scheme of $^{131}$Sn has been significantly expanded in the current work. Following the decay of the high-spin $21/2^+$ isomer of $^{131}$In, 13 new excited states and 24 new $\gamma$-ray transitions were observed. Following the decay of the low-spin $1/2^-$ isomer and $9/2^+$ ground states of $^{131}$In, 4 new excited states and 10 new $\gamma$-ray transitions were identified. This included a weak allowed $\beta$-transition from the $1/2^-$ isomer of $^{131}$In to the 3438-keV $(3/2^-)$ state in $^{131}$Sn expected to have a 1p-2h $\nu p_{3/2}(2h)$ configuration indicative of a robust $N=82$ shell closure in $^{131}$In.

Strong FF $\beta$-feeding of 27.8(7)\% was also observed from the $1/2^-$ state in $^{131}$In to the 332-keV $1/2^+$ state in $^{131}$Sn. This is in contrast to the 3.5(9)\% branching ratio reported in Ref.~\cite{PhysRevC.70.034312}. As the $1/2^-\rightarrow 1/2^+$ $\beta$ decay is used in the scaling of FF $\beta$-decay strengths~\cite{zhi13}, the significant revision in this $\beta$ branching ratio should be taken into account in future shell model calculations in the region. Furthermore, the newly observed excited states, $\beta$-decay branching ratios and $\gamma$-decay strengths should help guide future shell-model calculations of the nuclei in this region, including those required by $r$-process simulations.

The half-lives measured in this work were 323(55)~ms, 328(15)~ms and 265(8)~ms for the decays of the $21/2^+$, $1/2^-$ and $9/2^+$ states of $^{131}$In, respectively. Each of these half-lives is significantly longer than the half-life predicted from shell-model calculations~\cite{zhi13}. Further investigation into the discrepancy between the calculated and measured half-lives is therefore required.

\begin{acknowledgments}
This work has been partially supported by the Natural Sciences and Engineering Research Council of Canada (NSERC) and the Canada Research Chairs Program. The GRIFFIN spectrometer was funded by the Canada Foundation for Innovation, TRIUMF, and the University of Guelph. TRIUMF receives federal funding via a contribution agreement with the National Research Council of Canada.
\end{acknowledgments}

\bibliography{mybib}

\end{document}